\documentstyle[osa,manuscript,psfig]{revtex} 
\begin{document}
\title{Formation of color-singlet gluon-clusters and 
inelastic diffractive scattering}

\author{C. Boros\thanks{\emph{Present address:} 
                 Centre for Subatomic Structure of Matter (CSSM), 
                 University of Adelaide, Australia 5005.}, 
        Meng Ta-chung, 
        R. Rittel, 
        K. Tabelow,  
    and Zhang Yang\thanks{\emph{Present address:} 
                   Chinese Academy of Sciences, 
                   Institute of theoretical Physics, 
                   POB 2735, Beijing 100080, China.}\\
{\em Institut f\"ur Theoretische Physik, 
     FU Berlin, 
     14195 Berlin, 
     Germany}\\
{\em\small e-mail: meng@physik.fu-berlin.de}}
%

\maketitle

\begin{abstract}
This is the extensive follow-up report of a recent Letter in 
which the existence of self-organized criticality (SOC) in systems of
interacting soft gluons is proposed, and its consequences for
inelastic diffractive scattering processes are discussed.
It is pointed out, that color-singlet gluon-clusters can be formed
in hadrons as a consequence of SOC in
systems of interacting soft gluons, and that the properties of such 
spatiotemporal complexities can be probed experimentally by examing
inelastic diffractive scattering.
Theoretical arguments and experimental evidences supporting
the proposed picture are presented --- together with the result of a
systematic analysis of the existing
data for inelastic diffractive
scattering processes performed at different incident energies, and/or by
using different beam-particles. It is shown in particular 
that the size- and the
lifetime-distributions of such gluon-clusters can be directly
extracted from the data, and the obtained results exhibit universal
power-law behaviors --- in accordance with the expected
SOC-fingerprints.
As further consequences of SOC in systems of interacting soft gluons, the
$t$-dependence and the $(M_x^2/s)$-dependence of the double differential
cross-sections for inelastic diffractive 
scattering off proton-target are discussed. Here $t$ stands for the
four-momentum-transfer squared, $M_x$ for the missing mass, and $\sqrt{s}$ for
the total c.m.s. energy. 
It is shown, that the space-time properties of the color-singlet
gluon-clusters due to SOC, discussed above, lead to simple analytical 
formulae for $d^2\sigma/dt\,d(M_x^2/s)$  and  for $d\sigma/dt$, and
that the obtained results
are in good agreement with 
the existing data. Further experiments are suggested.
\end{abstract}
%


%
\section{Interacting soft gluons in the small-$x_B$ region of DIS}
\label{sec:1}
A number of striking phenomena have
been observed in recent 
deep-inelastic electron-proton scattering (DIS) experiments  
in the small-$x_B$ region. In particular it is seen, 
that the contribution of the gluons dominates\cite{r1},
and that large-rapidity-gap (LRG) events exist\cite{r2,r3,r3a}. 
The latter shows that the virtual photons in such processes may
encounter ``colorless objects'' originating from the proton.

The existence of LRG events in these and other\cite{r4,r5}
scattering processes
have attracted much attention, and
there has been much discussion \cite{r2,r3,r3a,r4,r5,r6,r7,r8,r9,r10}
on problems associated with the origin and/or the 
properties of such ``colorless objects''.
Reactions in which ``exchange'' of such ``colorless objects'' dominate
are known in the literature \cite{r3,r3a,r7,r8} as 
``diffractive scattering processes''.
While the concepts and methods used by different 
authors   in describing such processes
are in general very much different from one another,
all the authors
(experimentalists as well as theorists)
seem to agree on the following\cite{r8}
(see also Refs.
[\ref{r2}--\ref{r7}, \ref{r9}--\ref{r10a}]):
(a) Interacting soft gluons play a dominating role in
understanding the phenomena in the small-$x_B$ region of DIS in 
general, and in describing the properties of LRG events in particular.
(b) Perturbative QCD should be, and can be, used to describe the 
LRG events associated with high transverse-momentum ($p_\perp$)
jets which have been observed at HERA\cite{r9} and at the Tevatron\cite{r6}. 
Such events are, however, rather rare.
For the description of the bulk of LRG events, concepts 
and methods beyond the perturbative QCD, for example
Pomeron Models\cite{r7} based on Regge Phenomenology, are needed.
It has been suggested a long time ago 
(see the first two papers in Ref.[\ref{r7}])
that, in the QCD language,
``Pomeron-exchange'' can be interpreted as ``exchange of two or more
gluons'' and that such results can be obtained by calculating the
corresponding Feynman diagrams. It is generally felt that
non-perturbative methods should be useful in understanding ``the
small-$x_B$ phenomena'', but the
question, whether or how
perturbative QCD plays a role in such non-perturbative approaches does
not have an unique answer.

In a recent Letter \cite{r10a}, we proposed that the ``colorless
objects'' which play the dominating role in LRG events are
color-singlet gluon-clusters due to self-organized criticality, and
that optical-geometrical concepts and methods are useful
in examing the space-time properties of such objects.

The proposed picture \cite{r10a} is based on the following
observation: In a system of soft gluons
whose interactions are not negligible, 
gluons can be emitted and/or absorbed
at any time and everywhere in the system 
due to color-interactions between the members of the system as well as 
due to color-interactions of the members with
gluons and/or  quarks and antiquarks outside the system. In this
connection it is important to keep in mind that gluons interact
directly with gluons 
and that {\em the number of gluons in a system is not a conserved
quantity}. Furthermore, since in systems of interacting soft-gluons
the ``running-coupling-constant'' is in general greater than unity, 
non-perturbative methods are needed to describe the local
interactions associated with such systems. 
That is, such systems are in general extremely complicated, they are 
not only too complicated (at least for us) to take
the details of local interactions into account 
(for example by describing
the reaction mechanisms in terms of Feynman diagrams),
but also too complicated to apply well-known concepts
and methods in conventional Equilibrium Statistical Mechanics.
In fact, the accumulated
empirical facts about LRG events
and the basic properties of gluons prescribed by the QCD
are forcing us to accept the following
picture for such systems:

A system of interacting soft gluons can be, and should be considered as
{\it an open, dynamical, complex system with many degrees of freedom},
which is in general {\em far from equilibrium}.

In our search for an appropriate method to deal with such complex
systems, we are led to the following questions: Do we see comparable 
complex systems in Nature?
If yes, what
are the characteristic 
features of such systems, and what can we learn by studying such systems?

\section{Characteristic features of open dynamical complex systems}
\label{sec:2}
Open, dynamical, complex systems which are in general far from equilibrium
are {\em not} difficult to find in Nature ---
at least {\em not} in the macroscopic world! Such systems have been studied,
and in particular the following have been observed by
Bak, Tang and Wiesenfeld (BTW) some time ago\cite{r11}:
This kind of complex systems may evolve to 
self-organized critical states which lead to
fluctuations extending over all length- and
time-scales, and that 
such fluctuations manifest themselves in form of
spatial and temporal power-law scaling behaviors 
showing properties
associated with fractal
structure and flicker  noise, respectively. 

To be more precise, BTW\cite{r11} and many other
authors\cite{r12} proposed, and demonstrated by
numerical simulations, the following:  
Open, dynamical, complex systems of locally interacting objects which
are in general far from equilibrium
can evolve into self-organized
structures of states which are barely stable. A local perturbation of a
critical state may ``propagate'', in the sense that it spreads to (some)
nearest neighbors, and than to the next-nearest neighbors, and so on in
a ``domino effect'' over all length scales,  the size of 
such an ``avalanche'' can be as
large as the  entire
system. Such a ``domino effect'' eventually terminates after a total time $T$,
having reached a final amount of dissipative energy and having
effected a total spatial extension $S$. The quantity $S$ is called by
BTW the ``size'', and the quantity $T$ the ``lifetime'' of the 
avalanche --- named by BTW a ``cluster''
(hereafter referred to as BTW-cluster or BTW-avalanche). As  we
shall see in more details later on, it is of considerable importance to
note that a BTW-cluster {\it cannot}, and {\it  should not}
be identified with a cluster in the usual sense. 
It is an avalanche, 
{\it not} a {\it static} object 
with a fixed structure
which remains unchanged until it
decays after a time-interval (known as the lifetime in
the usual sense). 

In fact, it has been
shown\cite{r11,r12} that the 
distribution ($D_S$) of 
the ``size'' (which is a measure of
the dissipative energy, $S$) and the distribution 
($D_T$) of the lifetime
($T$) of BTW-clusters in  such open, 
dynamical, complex systems obey   power-laws:
\begin{equation}
\label{eq1}
D_S(S)\sim S^{-\mu},
\end{equation}
\begin{equation}
\label{eq2}
D_T(T)\sim T^{-\nu},
\end{equation}
where $\mu$ and $\nu$ are positive real constants. Such spatial and 
temporal power-law
scaling behaviors can be, and have been, considered
as the universal signals --- the ``fingerprints'' ---   of  the 
locally  perturbed
self-organized  critical states in such systems. 
It is
expected\cite{r11,r12} that the general concept of self-organized
criticality (SOC), which  is 
complementary to chaos, may be 
{\it the} underlying concept for temporal and spatial scaling in a wide class
of {\it open, non-equilibrium, complex systems} --- although it is not
yet known 
how the exponents of such power laws can be calculated analytically
from fundamental theories such that for gravitation or that for
electromagnetism. 

SOC has been {\em observed experimentally} in 
a large number of open, dynamical, complex systems in
non-equilibrium\cite{r11,r12,r14,r15,r16,r17} 
among which the following examples are
of particular interest, because they illuminate several aspects of
SOC which are relevant for the discussion in this paper.

First, the well-known Gutenberg-Richter law\cite{r13,r14}
for earthquakes as a special
case of Eq.(\ref{eq1}): 
In this case, earthquakes are BTW-clusters due to SOC. Here,
$S$ stands for the released energy (the magnitude)
of the observed earthquakes. $D_S(S)$ is the number of
earthquakes at which an energy $S$ is released.
Such a simple law is known to be valid
for all earthquakes, large (up to $8$ or $9$ in Richter scale)
or small! We note, the power-law behavior given by the
Gutenberg-Richter law implies in particular the following.
The question ``How large is a {\em typical} earthquake?'' does
not make sense!

Second, the sandpile experiments\cite{r11,r12} which show 
the simple regularities mentioned in Eqs.(\ref{eq1}) and (\ref{eq2}): 
In this example, we see how local perturbation can be caused by the 
addition of one grain of sand (note that we are dealing with 
an open system!). Here, 
we can also see how
the 
propagation of perturbation in form of ``domino effect'' 
takes place, and 
develops into BTW-clusters/avalanches of all possible sizes and durations.
The size- and duration-distributions are given by Eqs.(\ref{eq1}) 
and (\ref{eq2}), respectively.
This example is indeed a very attractive one,
not only because such 
{\em experiments} can be, and have been performed in {\em laboratories} 
\cite{r12}, but also because they can
be readily simulated on a PC\cite{r11,r12}.

Furthermore, it has been pointed out, and demonstrated
by simple models\cite{r12,r15,r16,r17},
that the concept of SOC can also be applied
to Biological
Sciences.
It is amazing to see how phenomena as complicated as Life 
and Evolution can be simulated
by simple models such as the ``Game of Life''\cite{r15} and
the ``Evolution Model''\cite{r16,r17}.

Having seen that systems of interacting soft-gluons
are open, dynamical, complex systems,
and that a wide class of open, dynamical, complex systems 
in the macroscopic world
evolve into self-organized critical states which lead to
fluctuations extending over all length- and time-scales,
it seems natural to ask the following:
Can such states and such fluctuations
also exist in the microscopic world --- on the
level of quarks and gluons? In particular: Can SOC be the dynamical
origin of
color-singlet gluon-clusters which play
the dominating role in inelastic diffractive scattering processes?

\section{SOC in inelastic diffractive scattering processes?} 
\label{sec:3}
One of the main goals of the present paper is
to answer the questions mentioned at the end of the last section.
We discuss in this and in the 
following four sections, {\em how to look for} signals of SOC in
systems of interacting gluons, and {\em what can we see} when we look
for signals of SOC in scattering processes in which systems of
interacting gluons play the dominating role.

Here, we explicitly see: (a) The fundamental properties of the
gluons, (b) the necessary conditions for the occurrence of SOC, and (c) 
the available technical possibilities strongly suggest that the most
favorable place to study the possible existence of such signals is (i) 
to look at the experimental results obtained in events associated with 
large rapidity gaps in deep-inelastic scattering, (ii) to look at the
experimental results in inelastic diffractive hadron-hadron
scattering, and (iii) to compare such observations with each other.

Having the special role played by ``the colorless objects''
in inelastic diffractive scattering in mind,
let us begin our discussion with the question: 
What {\em are} such ``colorless objects''? Up to now, we do not know
much about such objects. We know that they carry neither color- nor any
flavor-quantum numbers. We know that they exist in high-energy
reactions where soft gluons play the dominating role. We know that
they can be probed in diffractive scattering processes, in the sense
that they can interact with different beam-particles. But, there 
are a lot more which we do not know, for example: 
What is the mass of a typical
``colorless object''? What is the lifetime of a typical ``colorless
object''? Do such objects have distinct electromagnetic structures?
Are they hadron-like? Before more and better empirical facts about
such objects become available, guesses and/or speculations may be
helpful, provided that they agree with the existing data,
and they are consistent 
with the fundamental theoretical knowledge --- in particular
consistent with the
basic properties of the gluons (the direct gluon-gluon coupling
prescribed by the QCD-Lagrangian, the confinement, and the
non-conservation of gluon number, etc.). In this sense, we may wish to 
ask: Is it possible that the
colorless objects are BTW-clusters which exist due to SOC in systems
of interacting soft gluons? 
We are aware of the fact that the existence of SOC cannot (at least
cannot yet) be
derived from a basic theory such as QCD [Perhaps this can be and/or
should be compared with
the fact that the Gutenberg Richter law for earthquakes cannot (at least
cannot yet) be
derived from gravitational theory]. But, as in the case of earthquakes 
or any other open dynamical systems which leads to SOC, we can and we
should ask:
Can this be checked experimentally? Can this be done by looking for
characteristic properties of SOC --- in particular the
SOC-fingerprints mentioned in Eqs.(\ref{eq1}) and (\ref{eq2}) in the
relevant experiments?

To answer these questions,  
it is useful to recall the following: 
Since the ``colorless objects'' are color-singlets which can 
exist inside and/or outside the proton, the interactions between
such color-singlets as well as those between such objects and ``the
mother proton'' should be of Van der Waals type.
Hence, it is expected that
such a colorless object can be readily separated as an entire object
from the mother proton in 
scattering processes in which the momentum-transfer is sufficient 
to overcome the binding energy due to the Van der Waals type of
interactions. This means, in inelastic diffractive scattering 
the beam-particle (which is the virtual photon $\gamma^\star$ in DIS)
should have a chance to encounter one of the color-singlet
gluon-clusters. For the reasons mentioned above, the struck colorless
object can simply be ``knocked out'' and/or ``carried away'' by the
beam-particle in such a collision event. Hence, it seems that the
question whether ``the colorless objects'' are indeed BTW-clusters is
something that can be answered experimentally. In this connection we
recall that, according to the general theory of SOC\cite{r11,r12}, the 
size of a BTW-cluster is characterized by its dissipative energy, and
in case of systems of interacting soft gluons associated with the
proton, the dissipative energy carried by the BTW-cluster should be
proportional to the energy fraction ($x_P$) carried by the colorless
object. Hence, if the colorless object can indeed  be considered as a
BTW-cluster due to SOC, we should be able to obtain information about
the size-distribution of such color-singlet gluon-clusters by examing
the $x_P$-distributions of LRG events in the small-$x_B$ region of DIS.   

\begin{figure}
\psfig{figure=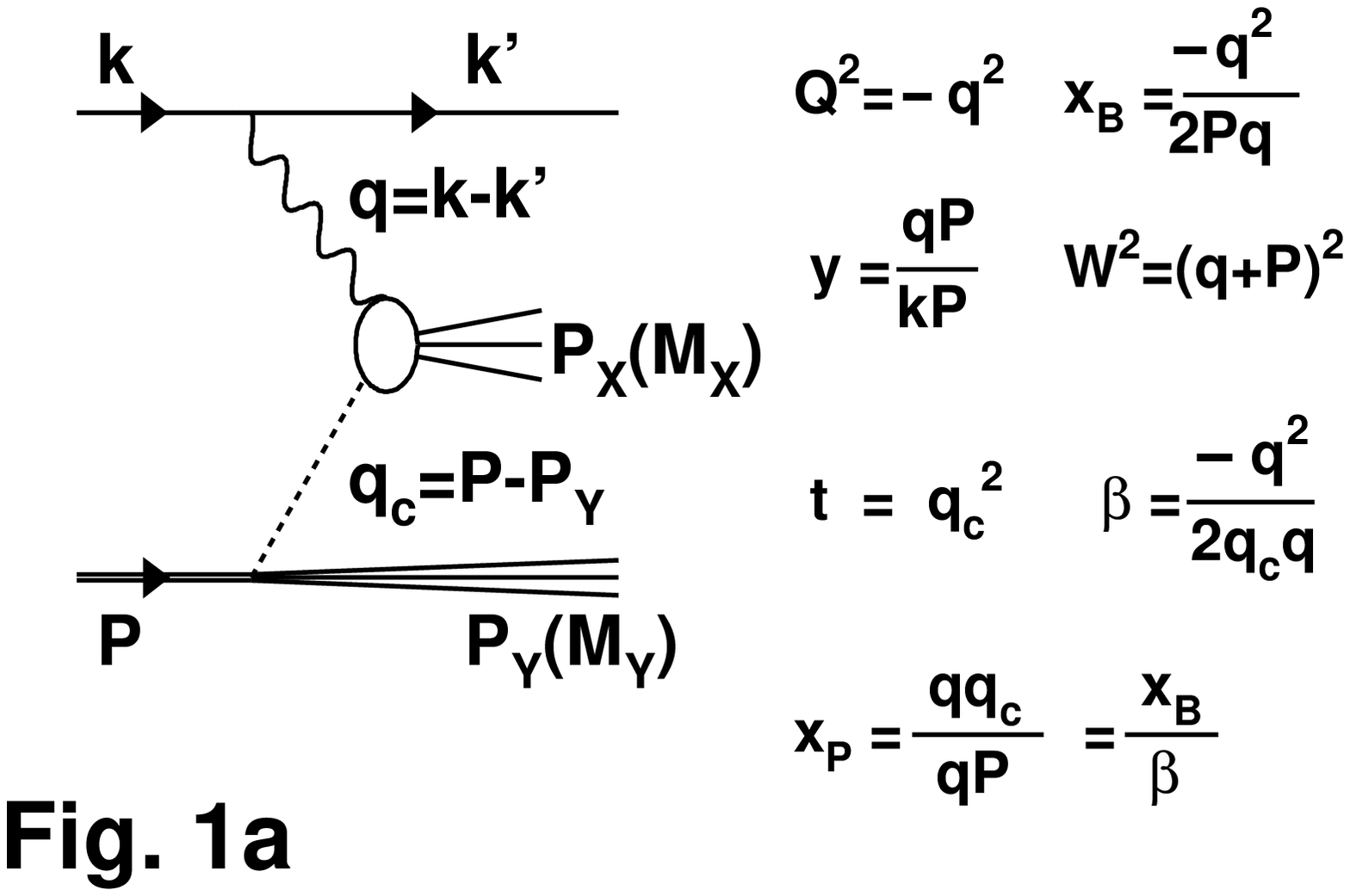,width=16.0cm}
\psfig{figure=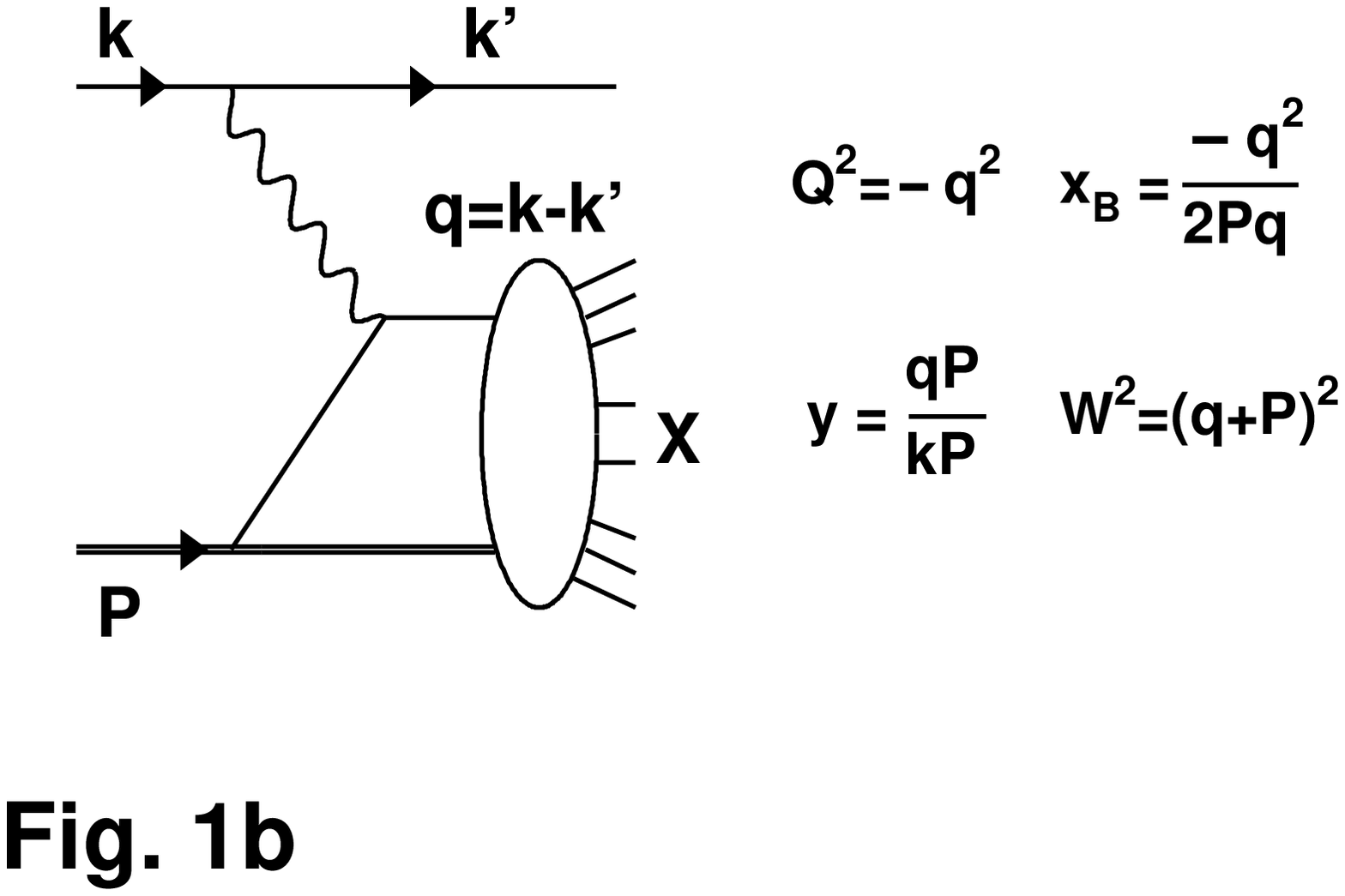,width=16.0cm}
\caption{\label{figure1}
The well-known Feynman diagrams (a) for diffractive and (b) for normal
deep-inelastic 
electron-proton scattering are shown together with 
the relevant kinematical variables which describe
such processes.}
\end{figure}

Having this in mind, we now take a closer look at 
the measured \cite{r3}
``diffractive structure function'' 
\mbox{$F_2^{D(3)}(\beta,Q^2;x_P)\equiv \int dt F_2^{D(4)}(\beta,Q^2;x_P,t)$}.
Here, we note that $F_2^{D(4)}(\beta,Q^2;x_P,t)$ 
is related \cite{r3,r3a,r7,r8,r9} to the 
differential cross-section for large-rapidity-gap
events 
\begin{equation}
\label{eq3}
{d^4\sigma^D\over d\beta dQ^2 dx_P dt}={4\pi\alpha^2\over\beta
Q^4}(1-y+{y^2\over 2})F_2^{D(4)}(\beta,Q^2;x_P,t),
\end{equation}
in analogy to 
the relationship between the corresponding quantities
[namely $d^2\sigma/(dx_B\,dQ^2)$ and $F_2(x_B,Q^2)$]
for normal deep-inelastic electron-proton scattering events
\begin{equation}
\label{eq4}
{d^2\sigma\over dx_BdQ^2}={4\pi\alpha^2\over
x_BQ^4}(1-y+{y^2\over 2})F_2(x_B,Q^2).
\end{equation}
The kinematical variables, in particular $\beta$, $Q^2$, $x_P$ and $x_B$ 
(in both cases) are directly measurable quantities, the  definitions 
of which are shown in Fig.\ref{figure1} together with the corresponding
diagrams of the
scattering processes. We note
that, although these variables are 
Lorentz-invariants, it is sometimes convenient to interpret them in a
``fast moving frame'', for example the electron-proton center-of-mass 
frame  where the proton's 3-momentum $\vec P$ is large (i.e. its 
magnitude $|\vec P|$ and thus the energy $P^0\equiv (|\vec P|^2+M^2)^{1/2}$ 
is much larger than the proton mass $M$). While $Q^2$ characterizes 
the virtuality of the space-like photon
$\gamma^\star$, $x_B$ can be interpreted, 
in  such a ``fast moving frame'' (in the framework 
of the celebrated parton model),  as the
fraction of proton's energy $P^0$ (or longitudinal momentum $|\vec P|$)
carried by the struck charged constituent. 

We recall, in the framework 
of the  parton model, $F_2(x_B,$ $Q^2)/x_B$ for ``normal events''
can be interpreted  as the sum of the probability densities 
for the above-mentioned $\gamma^\star$ to interact with 
a  charged constituent of the proton. In analogy to this, 
the quantity
$F_2^{D(3)}(\beta,Q^2;x_P)/\beta$ for LRG events
can be interpreted  as the sum of the probability 
densities for $\gamma^\star$ to interact with 
a charged constituent which 
carries a fraction $\beta\equiv x_B/x_P$ of the energy (or longitudinal 
momentum) of  the colorless object,
under the  condition that the colorless object
(which we associate with a system of interacting soft gluons) carries a
fraction $x_P$
of proton's energy (or longitudinal momentum). 
We hereafter denote this
charged-neutral and color-neutral gluon-system by
$c^\star_0$ (in Regge pole models\cite{r7} this object  is known as 
the ``Pomeron''). 
Hence, by comparing Eq.(\ref{eq3}) with Eq.(\ref{eq4}) 
and by comparing the two 
diagrams shown in Fig.\ref{figure1}a and Fig.\ref{figure1}b, 
it is tempting to draw the following conclusions: 

The diffractive process is nothing else but
a process in which the virtual photon $\gamma^\star$ 
encounters a $c_0^\star$,
and $\beta$ is nothing else but the Bjorken-variable with respect to 
$c_0^\star$ (this is why it is called $x_{BC}$ in Ref.[\ref{r10}]). 
This means, 
a diffractive $e^-p$ scattering event can be envisaged as an event in 
which the virtual photon $\gamma^\star$ collides with ``a $c_0^\star$-target'' 
instead of ``the  proton-target''. 
Furthermore, since  $c_0^\star$ is charge-neutral,
and  a  photon can only directly interact with an object
which has electric charges and/or magnetic moments,
it is tempting to assign $c_0^\star$ an
electromagnetic structure function $F_2^{c}(\beta, Q^2)$,
and study the interactions between the virtual photon and the quark(s)
and antiquark(s) inside $c_0^\star$.
In such a picture
(which should be formally the same as that of
Regge pole models\cite{r7},
if we would replace the $c_0^\star$'s by ``pomerons'')
we are confronted with the following two questions: 

First, is it possible and meaningful to discuss the $x_P$-distributions of 
the $c_0^\star$'s without knowing the intrinsic properties, in particular the 
electromagnetic structures, of such objects? 

Second, are the $c_0^\star$'s hadron-like, such that the electromagnetic 
structure of a (a typical, or an average) $c_0^\star$ can be studied 
in the same way as those for ordinary hadron?

Since we wish to begin the quantitative discussion with
something familiar to most of the
readers in this community, and we wish
 to differentiate between the
conventional-approach and the SOC-approach, we would like to 
discuss the second question here, and leave the first question to
the next section. 
We recall (see in particular the last two papers in Ref.[\ref{r7}]),
in order to see whether
the second question
can be answered
in the {\em affirmative}, 
we need to know {\em whether}
$F_2^{D(3)}(\beta,Q^2;x_P)$ can be factorized in the form
\begin{equation}
\label{eq5}
F_2^{D(3)}(\beta, Q^2;x_P)=f_c(x_P)F_2^c(\beta,Q^2).
\end{equation}
Here, $f_c(x_P)$ plays the role of a ``kinematical factor''
associated with the ``target $c_0^\star$'', 
and $x_P$ is  the fraction
of proton's energy (or longitudinal momentum) carried by
$c_0^\star$. [We could call $f_c(x_P)$
``the $c_0^\star$-flux'' ---  in exactly the same 
manner  as in Regge pole models\cite{r7}, where it is called 
``the pomeron flux''.] $F_2^c(\beta,Q^2)$ is 
``the electromagnetic structure function of $c_0^\star$''
[the counterpart of $F_2(x_B,Q^2)$ of the proton] which
--- in analogy to proton (or any other hadron) ---
can be expressed as
\begin{equation}
\label{eq6}
\frac{F_2^c(\beta,Q^2)}{\beta}
= \sum_i e_i^2 [q_i^c(\beta,Q^2)+\bar q_i^c(\beta,Q^2)],
\end{equation}
where $q_i^c(\bar q_i^c)$ stands for the probability 
density for $\gamma^\star$ 
to interact with a quark (antiquark) of flavor $i$ and electric
charge $e_i$ which carries a fraction $\beta$ of the energy 
(or longitudinal momentum)
of $c_0^\star$. It is clear that 
Eq.(\ref{eq6}) should be valid for all $x_P$-values in this kinematical
region, that is, both the right- and the left-hand-side
of Eq.(\ref{eq6}) should be independent of the energy (momentum) carried
by the ``hadron'' $c_0^\star$.

\begin{figure}
\psfig{figure=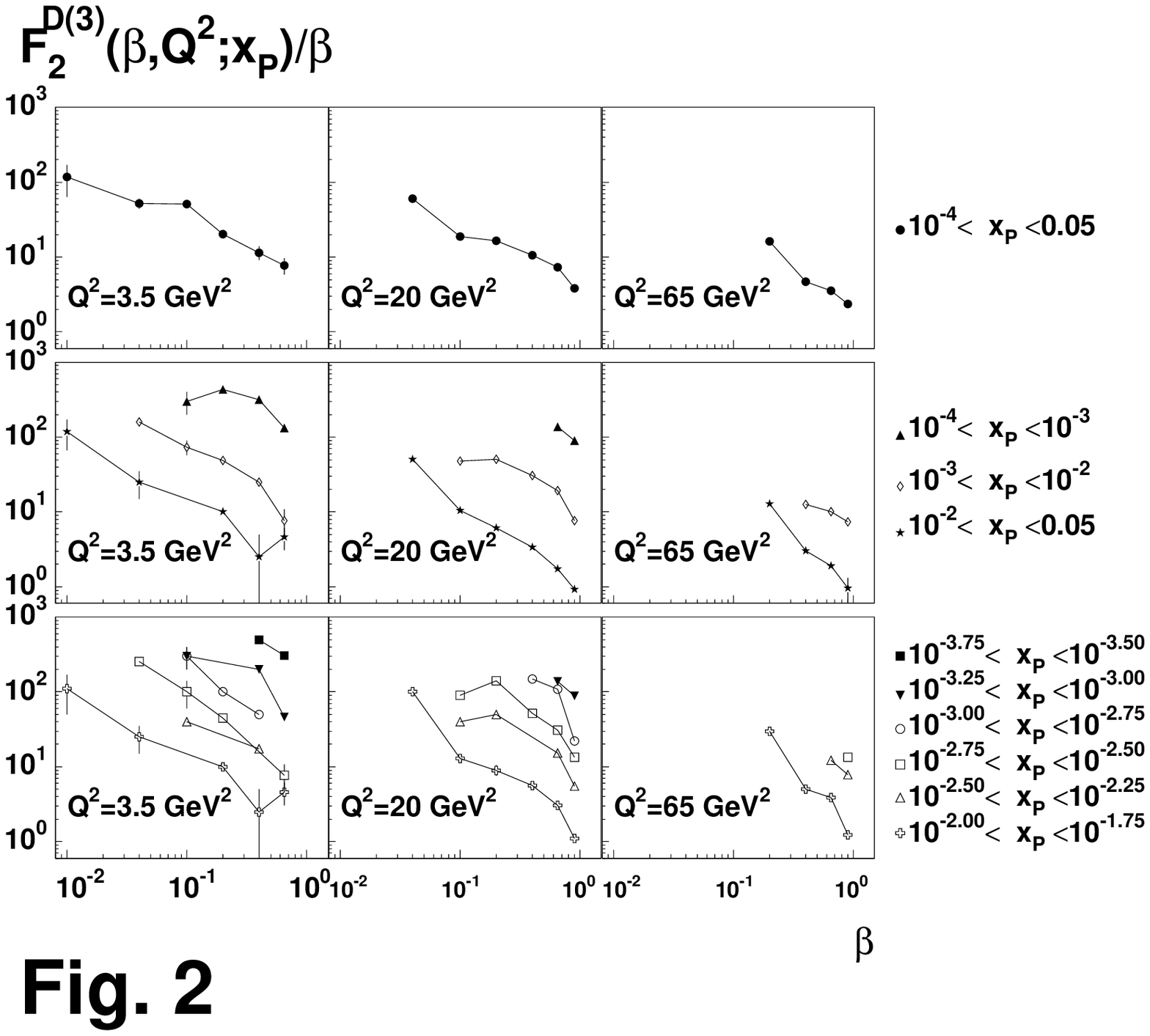,width=16.0cm}
\caption{\label{figure2}
$F_2^{D(3)}(\beta,Q^2;x_P)/\beta$ is plotted as a function of 
$\beta$ for given $x_P$-intervals and for fixed $Q^2$-values. 
The data are taken from Ref.[\ref{r3}].
The  lines  are only to guide the eye. }
\end{figure}

Hence, to find out experimentally whether the second question can be 
answered in the affirmative, we only need to check whether the
data are in agreement with the assumption
that $F_2^c(\beta , Q^2)$ prescribed by Eqs.(\ref{eq5}) and
(\ref{eq6}) exists. For such a test,
we take the existing
data\cite{r3} and plot $\log [F_2^{D(3)}(\beta, Q^2;x_P)/\beta]$ 
against $\log\beta$ for different $x_P$-values.
We note, under the assumption 
that the factorization shown in Eq.(\ref{eq5})
is valid, the $\beta$-dependence for a given $Q^2$ in
such a plot should have exactly the same form as that in the  
corresponding
$\log [F_2^{c}(\beta, Q^2)/\beta]$ vs $\log \beta$ plot;
and that the latter is the analog of
$\log [F_2(x_B, Q^2)/x_B]$ vs $\log x_B$ plot for normal events. 
In Fig.\ref{figure2} we show the result of such
plots for three fixed $Q^2$-values (3.5, 20 and 65 GeV$^2$,
as representatives of three different ranges in $Q^2$). 
Our goal is to examine whether or
how the $\beta$-dependence of the function given in
Eq.(\ref{eq6}) changes with $x_P$. In principle,
if there were enough data points, we should, and we could, do such
a plot for the data-sets associated with every $x_P$-value.
But, unfortunately there are not so much data at present.
What we can do, however, is to consider
the $\beta$-distributions in different $x_P$-bins, and to vary
the bin-size of $x_P$,
so that we can explicitly 
see whether/how the shapes of the $\beta$-distributions
change. The results are shown  
in Fig.\ref{figure2}. The $\beta$-distribution in the first 
row, corresponds to the integrated value $\tilde{F}^D_2(\beta, Q^2)$
shown in the literature\cite{r3,r8}.
Those in the second and in the third row are obtained by considering
different bins and/or by
varying the sizes of the bins.
By joining the points associated with a given $x_P$-interval
in a plot for a given $Q^2$,
we obtain the $\beta$-distribution for a $c_0^\star$ carrying 
approximately the amount of energy $x_P P^0$, encountered 
by a photon of virtuality $Q^2$. Taken together with Eq.(\ref{eq6}) we can 
then extract the distributions $q_i^c(\beta, Q^2)$ and
 $\bar{q}_i^c(\beta, Q^2)$  for this $Q^2$-value, provided
that $F_2^c(\beta, Q^2)/\beta$ is independent of $x_P$.
But, as we can see in Fig.\ref{figure2}, the existing data\cite{r3}
show that the $x_P$-dependence of this function is far from 
being negligible!
Note in particular 
that according to Eq.(\ref{eq5}), by choosing a suitable $f_P(x_P)$ 
we can shift the curves for different $x_P$-values in the vertical 
direction (in this log-log plot); but {\em we can never change 
the shapes of the $\beta$-distributions} which are different for
different $x_P$-values!  

In order to see, and to realize, the meaning of the $x_P$-dependence
of the distributions of the charged constituents of $c^\star_0$ 
expressed in terms of  $F_2^c(\beta, Q^2)/\beta$ 
in LRG events [see Eqs.(\ref{eq5}) and (\ref{eq6})], 
let us, for a moment, consider
normal deep-inelastic scattering events in the 
$x_B$-region where quarks dominate ($x_B > 0.1$, say).
Here we can plot the data for 
$\log [F_2(x_B, Q^2)/x_B]$ as a function of $\log x_B$ obtained
at {\em different incident energies ($P^0$'s)} of the proton.
{\em Suppose} we see, that 
at a given $Q^2$, the data for $x_B$-distributions taken
at different values
of $P^0$ are very much different.
{\em Would} it still be possible to introduce $F_2(x_B,Q^2)$
as ``the electromagnetic  structure function'' of the proton,
from which we can extract the $x_B$-distribution of the quarks
$q_i(x_B,Q^2)$ at a given $Q^2$?
The fact that it is not possible to assign
an $x_P$-independent
structure function $F_2^c(\beta, Q^2)/\beta$ to $c_0^\star$ which
stands
for the ``Pomeron'', and whose ``flux'' $f_c(x_P)$
is expected to be independent of $\beta$ and $Q^2$,
deserves to be taken seriously.
It strongly suggests that the following picture
{\em cannot} be true:
``There exists an universal colorless object 
(call it Pomeron or $c_0^\star$ or something else)
the exchange of which describes diffractive scattering 
in general and DIS off proton in particular. This object is
hadron-like
in the sense that it has not only a typical size and a typical
lifetime, but also a typical electromagnetic structure which can
e.g. be measured and described by an electromagnetic structure
function''.

In summary of this section, we note that 
the empirical facts mentioned above show that {\em no}
energy-independent electromagnetic structure function can be assigned
to the expected universal
colorless object $c_0^\star$. 
This experimental fact is of considerable importance, because 
it is the first indication that, if there is {\em an universal} ``colorless
object'', this object {\em cannot} be considered as 
an ordinary hadron. It has to be
something else! In fact, as we shall see below, this
property is closely related to the observation
that such an
object {\em cannot} have a typical size, or a typical
lifetime.
To be more precise, the fact that the data \cite{r3} {\em cannot}
accommodate the simple factorization assumption shown in Eq.(\ref{eq5}), in
which a universal Pomeron flux with a unique hadron-like Pomeron
structure function exists, can be considered as further support to the 
proposed SOC picture because a BTW-cluster, which has neither a
typical size nor a typical lifetime, {\em cannot} have an universal
static structure. With these characteristic properties of the
colorless objects in mind, we see \cite{r17a} an overlap between the
SOC-picture and the partonic picture for Pomeron and/or Pomeron and
Reggeon \cite{r7,r17a} in which, beside the Pomeron, exchange of ( in
general infinite number of) subleading trajectories are possible. In
fact, it has been reported \cite{r3,r3a} that very good agreement can be
achieved between the data \cite{r3,r3a} and these types of models. Hence, in 
order to differentiate between the two approaches, it is not only
useful but also necessary to examine the corresponding predictions for 
the dependence on the invariant momentum transfer $t$. This will be
discussed in detail in Sections \ref{sec:8} - \ref{sec:11}. 
\section{Distributions of the  gluon-clusters}
\label{sec:4}
After having seen that the existing data 
does not allow us to assign an energy-independent electromagnetic
structure function to ``the colorless object'' such that the universal 
colorless object ($c_0^\star$) can be treated as an ordinary hadron,
let us now come back 
to the first question in Section \ref{sec:3}, 
and try to find out whether it is
never-the-less possible, and meaningful, to talk about the 
$x_P$-distribution of $c_0^\star$. 
As we shall see in this section,
the answer to this question is Yes! 
Furthermore, we shall also see,
in order to answer this question
in the affirmative, 
we {\em do not} need the factorization mentioned 
in Eq.(\ref{eq5}), 
and we {\em do not} need to know whether the $c_0^\star$'s are
hadron-like. But, as we have already mentioned above, it is 
of considerable importance 
to discuss the second question 
so that we can understand the origin and 
the nature of the $c_0^\star$'s. 

In view of the fact that we do use the concept ``distributions
of gluons'' 
in deep-inelastic lepton-hadron scattering, although the gluons  
do not directly interact with the virtual photons,
we  shall try to introduce the notion ``distribution of 
$c_0^\star$'' in a similar manner. 
In order to see what we should do for the introduction
of such distributions, let us recall the following:

For normal deep-inelastic $e^-p$ collision
events, the structure function $F_2(x_B, Q^2)$ can be expressed
in terms of the distributions of partons, where the partons are 
not only quarks and antiquarks, but also gluons which 
can contribute to the structure function by quark-antiquark 
pair creation and annihilation. 
In fact, in order to satisfy energy-momentum-conservation
(in the electron-proton system),
the contribution of the gluons $x_gg(x_g,Q^2)$ has to be taken into account
in the energy-momentum sum rule
for all measured $Q^2$-values. Here, we denote by 
$g(x_g,Q^2)$ the probability density 
for the virtual photon $\gamma^\star$ (with virtuality $Q^2$) to meet a
gluon which carries the energy (momentum) fraction $x_g$ of the proton,
analogous to $q_i(x_B, Q^2)$ [or $\bar q_i(x_B, Q^2)$] which 
stands for the probability density for this $\gamma^\star$ 
to interact with a quark (or an antiquark) of flavor $i$ and electric 
charge $e_i$ which carries the energy (momentum) fraction $x_B$ of the 
proton. We note, while both $x_B$ and $x_g$ stand for energy 
(or longitudinal momentum) fractions carried by partons, 
the former can be, but the latter {\em cannot} be directly
measured. 

Having these, in particular the energy-momentum sum rule in mind,
we immediately see the following: In a given
kinematical region
in which the contributions of only
one category of partons (for example quarks for $x_B > 0.1$ or
gluons for $x_B < 10^{-2}$) dominate, the structure
function $F_2(x_B,Q^2)$ can approximately
be related to the 
distributions of that particular kind of partons in a
very simply manner. In fact,
the expressions below can be, and have been,
interpreted as the probability-densities for the virtual photon $\gamma^\star$
(with virtuality $Q^2$) to meet a quark or a gluon which carries the energy
(momentum) fraction $x_B$ or $x_g$ respectively. 

\begin{eqnarray}
\label{eq7}
{F_2(x_B,Q^2)\over x_B}&\approx& \sum_i e_i^2\, q_i(x_B,Q^2) 
\mbox{\hspace*{1cm}or\hspace*{1cm}} \nonumber\\
{F_2(x_B,Q^2)\over x_g}&\approx& g(x_g,Q^2)\mbox{\ .}
\end{eqnarray}
The relationship between $q_i(x_B,Q^2)$,
$g(x_g,Q^2)$ and
$F_2(x_B, Q^2)$ as they stand  in Eq.(\ref{eq7})
are general
and formal (this is the case especially for that between $g$ and
$F_2$) in the following sense: 
Both $q_i(x_B, Q^2)$ and $g(x_g,Q^2)$ contribute to the
energy-momentum sum rule and both of them are in accordance
with 
the assumption that partons 
of a given category
(quarks or gluons)
dominate a given kinematical region
(here $x_B>0.1$ and $x_B<10^{-2}$, respectively).
But, neither the dynamics which leads to the observed $Q^2$-dependence
nor the relationship between $x_g$ and $x_B$ are given. This means,
{\it without further theoretical inputs}, the simple expression for
$g(x_g, Q^2)$ as given by Eq.(\ref{eq7}) is {\it practically useless}!

Having learned this, we now discuss what happens
if we assume, in diffractive lepton-nucleon scattering, that
the color-singlet gluon-clusters ($c_0^\star$'s) dominate the 
small-$x_B$ region ($x_B< 10^{-2}$, say). In this simple picture, we
are assuming that the following is approximately true: 
The gluons in this region appear predominantly in form
of $c_0^\star$'s. The interaction
between  the struck $c_0^\star$
and the rest of the proton
can be neglected during the 
$\gamma$-$c_0^\star$ collision such that
we can apply impulse-approximation to the $c_0^\star$'s in this
kinematical region.  
That is, here we can
introduce 
--- in the same manner as we do for
other partons
(see Eq.(\ref{eq7})), a
probability density $D_S(x_P|\beta,Q^2)$ for $\gamma^\star$ in the 
diffractive scattering process to ``meet'' a $c_0^\star$ which carries 
the fraction $x_P$ of the proton's 
energy $P^0=(|\vec{P}|^2+M^2)^{1/2} \approx |\vec{P}|$
(where $\vec{P}$ is the momentum and $M$ is the mass of the proton).
In other words, 
in 
diffractive scattering events
for processes in the  kinematical region
$x_B < 10^{-2}$, we should have, instead of $g(x_g,Q^2)$, the following:
\begin{equation}
\label{eq8}
{F_2^{D(3)}(\beta,Q^2;x_P)\over x_P}\approx D_S(x_P|\beta,Q^2)\,.
\end{equation}
Here, $x_PP^0$ is the energy carried by $c_0^\star$, 
and $\beta$ indicates the corresponding fraction carried by
the struck charged constituent in $c_0^\star$.
In connection with the similarities and the differences between 
$q_i(x_B,Q^2)$, $g(x_B,Q^2)$ in Eq.(\ref{eq7}) and $D_S(x_P|\beta, Q^2)$
in Eq.(\ref{eq8}), it is
useful to note in particular the significant difference 
between $x_g$ and $x_P$, 
and thus that between 
the $x_g$-distribution $g(x_g,Q^2)$ of the gluons and the
$x_P$-distribution $D_S(x_P|\beta, Q^2)$ of the $c^\star_0$'s: Both $x_g$ and 
$x_P$ are energy (longitudinal momentum) fractions of charge-neutral 
objects, with which
$\gamma^\star$ {\it cannot} directly interact. But, in contrast to $x_g$,
$x_P$ {\it can be directly measured in experiments}, namely by 
making use of the kinematical relation
\begin{equation}
\label{eq9}
x_P\approx {Q^2+M_x^2\over Q^2+W^2},
\end{equation}
and
by measuring the quantities $Q^2$, $M_x^2$ and $W^2$
in every collision event. Here, $Q$, $M_x$
and $W$ stand respectively for the invariant momentum-transfer from
the incident electron, the invariant-mass of the final hadronic state
after the $\gamma^\star-c_0^\star$ collision, and the invariant mass of the
entire hadronic system in the collision between $\gamma^\star$ and the
proton. Note that $x_B\equiv\beta x_P$, hence $\beta$ is also
measurable. This means, in sharp contrast to $g(x_g,Q^2)$, {\it
experimental information} on $D_S(x_P|\beta, Q^2)$ 
in particular its $x_P$-dependence
can be obtained ---
{\it without further theoretical inputs}!
 
\section{The first SOC-fingerprint: Spatial scaling}
\label{sec:5}

We mentioned at the beginning of Section \ref{sec:3}, that in order
to find out whether the concept of SOC
indeed plays a role in diffractive DIS we need to check the
fingerprints of SOC shown in Section \ref{sec:2}, and that such tests 
can be made by examing the 
corresponding cluster-distributions obtained from experimental data.
We are now ready to do this, because we have
learned in Sections \ref{sec:3} and \ref{sec:4}, that it is not only
meaningful but also possible to extract $x_P$-distributions from the 
measured diffractive structure functions,
although the $c_0^\star$'s {\em cannot} be treated as hadrons.
In fact, as we can explicitly see
in Eqs.(\ref{eq8}) and (\ref{eq9}), 
in order to extract the $x_P$-dependence of
the $c_0^\star$'s from the data, detailed knowledge about the intrinsic
structure of the $c_0^\star$'s are not necessary.

\begin{figure}
\psfig{figure=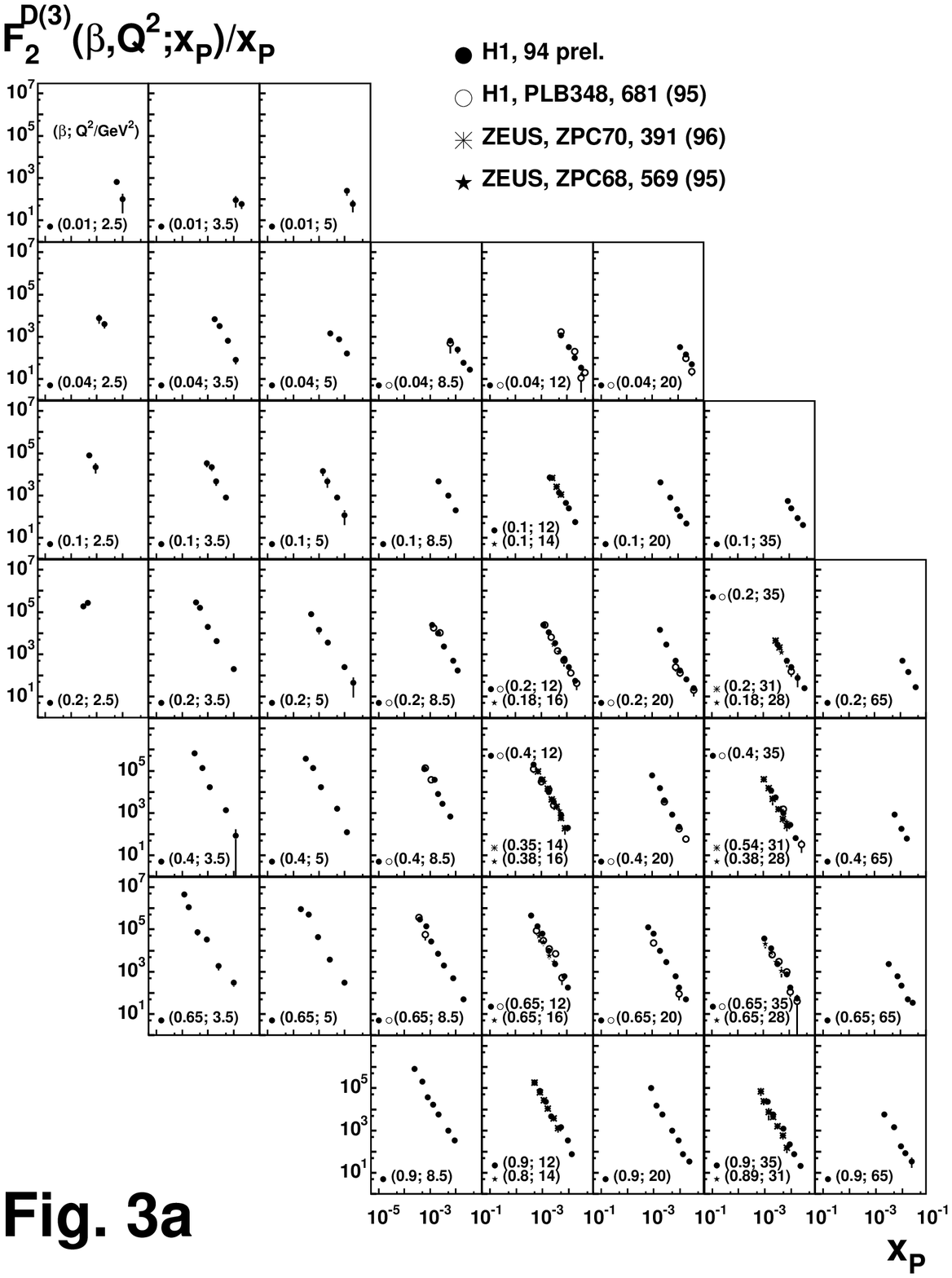,width=16.0cm}
\psfig{figure=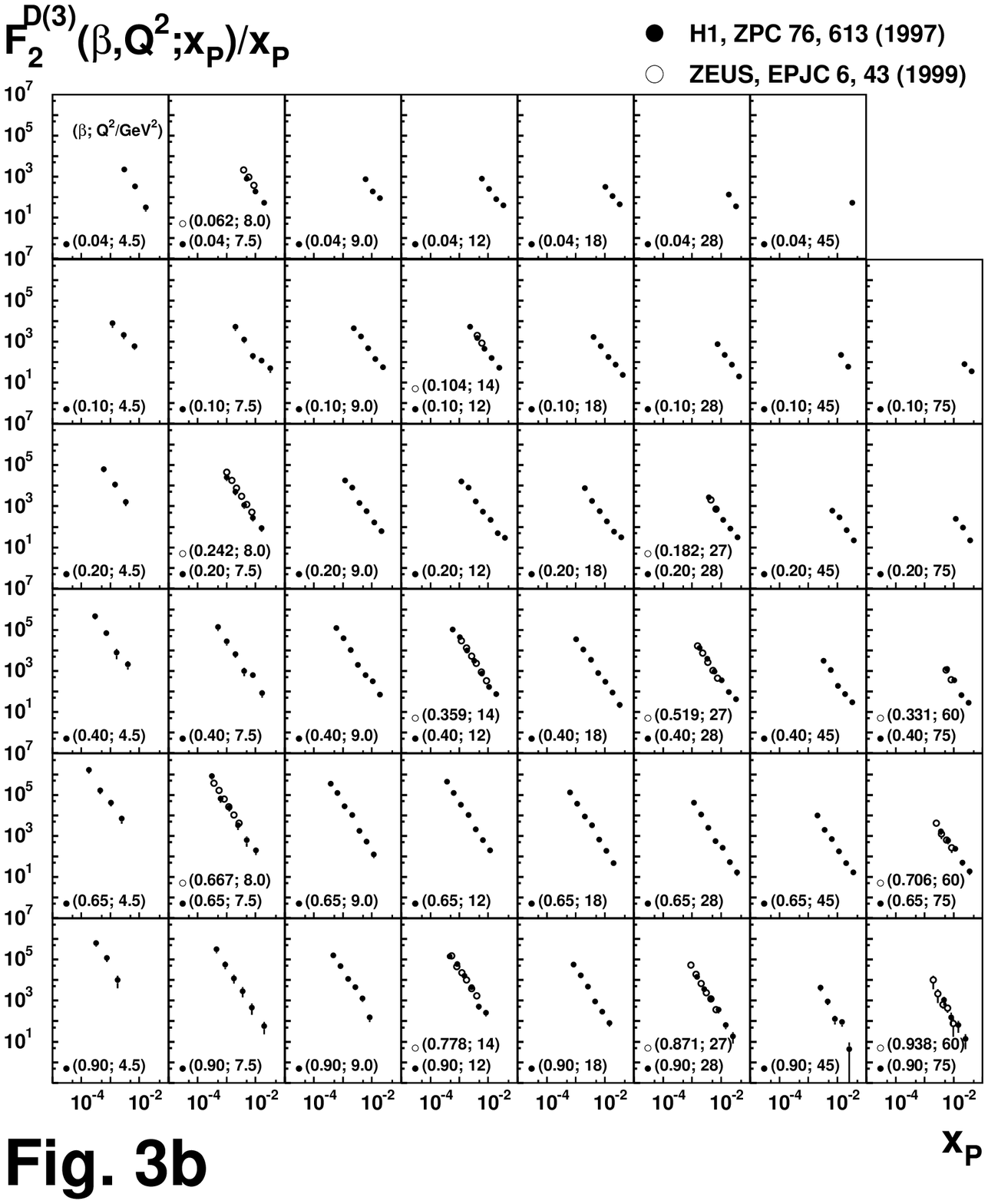,width=16.0cm}
\caption{\label{figure3} 
(a) $F_2^{D(3)}(\beta,Q^2;x_P)/x_P$ is plotted as a function of
$x_P$  for different values of $\beta$ and $Q^2$.  
The data are taken from 
Ref.[\ref{r3}]. (b) same as (a) but with data taken from
Ref.[\ref{r3a}]. Note that in these log-log plots, almost all existing
data points lie on straight-lines with approximately the {\em same}
slope irrespective of the values of $Q^2$ and/or $\beta$.}
\end{figure}

Having these in mind, we now
consider $D_S$ as a function of $x_P$ for given values
of $\beta$ and $Q^2$,
and plot 
$F_2^{D(3)}(\beta,Q^2;x_P)/x_P$ against $x_P$
for different sets of $\beta$ and $Q^2$. The results of such  
log-log plots are shown in Fig.\ref{figure3}.  As we can see,
the data\cite{r3,r3a} suggest
that the probability-density for the virtual photon $\gamma^\star$  to 
meet a color-neutral and charged-neutral object $c_0^\star$ with energy
(longitudinal momentum) fraction $x_P$ has a power-law behavior in
$x_P$, and the exponent of this power-law
depends very little on $Q^2$ and $\beta$.  
This is to be compared with $D_S(S)$ in Eq.(~\ref{eq1}), where $S$,
the dissipative energy (the size of the BTW-cluster)
corresponds to the energy of the
system $c_0^\star$. The latter is $x_PP^0$, where $P^0$ is the
total energy of the proton.

It means, the existing data\cite{r3,r3a} show that
$D_S(x_P | \beta, Q^2)$ exhibits the same kind of power-law behavior
as the size-distribution of BTW-clusters.
This result is 
in accordance with the expectation that
self-organized critical phenomena may exist
in systems of interacting soft gluons in
diffractive deep-inelastic electron-proton 
scattering processes.

We note, up to now, we have only argued (in Section \ref{sec:1}) that
such gluon-systems are
open, dynamical, complex systems
in which SOC may occur, and we have mentioned (in Section \ref{sec:2}) the 
ubiquitousness of SOC in Nature.
Having seen the experimental evidence
that one of the ``SOC-fingerprints'' (which are 
necessary conditions for the existence of
SOC) indeed exists, let us now take a second look at such
gluon-systems from a theoretical standpoint.
Viewed from a 
``fast moving frame'' which can
for example be the electron-proton c.m.s. frame,  
such systems
of interacting soft gluons are part of the proton
(although color-singlets can also be outside the confinement
region). Soft gluons can be
intermittently emitted or absorbed by gluons in such a system, 
as well as by gluons, 
quarks and antiquarks outside the system. 
The emission- and absorption-processes are due to local interactions
prescribed by the well-known QCD-Lagrangian (here ``the running
coupling constants'' are in general large,
because the distances between the interacting colored objects
cannot be considered as ``short''; remember that the 
spatial dimension of a $c_0^\star$ can be much
larger than that of a hadron!). 
In this connection, it is useful to keep the following in mind:
Due to the complexity of the system,
details about the local interactions may be relatively
unimportant, while
general and/or global features --- for example 
energy-flow between different parts (neighbors and neighbor's
neighbors $\ldots$) of the system --- 
may play an important role. 

How far can one go in neglecting dynamical details when one 
deals with such open 
complex systems? In order to see this, let us
recall how Bak and Sneppen\cite{r16}
succeeded in modeling
some of the essential aspects of 
The Evolution in Nature.
They consider the ``fitness'' of different ``species'', related to one
another through a ``food chain'', and assumed
that the species with the lowest fitness
is most likely to disappear or mutate at the next time-step
in their computer simulations. 
The crucial step in their simulations
that {\em drives} evolution is the adaption of the individual species to
its present {\em environment} (neighborhood) through mutation 
and selection of a 
fitter variant.
Other interacting species form part of the {\em environment}.
This means, the neighbors will be influenced by
every time-step.
The result these authors
obtained strongly suggests
that the process of evolution is
a self-organized critical phenomenon. One of the essential
simplifications they made in their evolution models\cite{r16,r17}
is the following: Instead of the explicit 
connection between
the fitness and the configuration of the
genetic codes,
they use random numbers for the fitness of the
species. 
Furthermore, as they have pointed out in their papers, they
 could in principle have chosen to model evolution on a less
coarse-grained scale by considering mutations at the individual
level rather than on the level of species, but that would make the 
computation prohibitively difficult.

Having these in mind, we are naturally led to the questions: 
Can we consider the creation and
annihilation processes of colorless
systems of interacting soft gluons associated
with a proton as ``evolution'' in a microscopic world?
Before we try to build models for a quantitative description
of the data, can we simply apply the existing evolution
models\cite{r16,r17} to such open, dynamical, complex 
systems of interacting soft gluons,
and check whether some of the essential features 
of such systems
can be
reproduced?

To answer these questions, we now report on the result of our 
first trial in this direction: 
Based on the fact that we know {\em very little} about
the detailed reaction mechanisms in such gluon-systems and 
{\em practically}
{\em nothing} about their structures, we simply {\em ignore} them, 
and assume that they are self-similar in space
(this means, color-singlet gluon-clusters ($c_0^\star$) 
can be considered as clusters of $c_0^\star$'s and so on). Next,
we divide them into an arbitrary given number of subsystems $s_i$
(which may or may not have the same size). Such a system is open,
in the sense that neither its energy $\varepsilon_i$, nor
its gluon-number $n_i$ has a fixed value. Since we do not
know, in particular, how large the $\varepsilon_i$'s are, we  
use random numbers. As far as the $n_i$'s are concerned, since
we do not know how these numbers are associated with the energies
in the subsystems $s_i$, except that they are not conserved
quantities,
we just ignore them, and consider only the $\varepsilon_i$'s.
As in Ref.[\ref{r16}] or in Ref.[\ref{r17}], the random number of this
subsystem as well as those of the fixed\cite{r16} or random (see the
first paper of Ref.[\ref{r17}]) neighbors will be changed at every time-step.
Note, this is how we simulate the processes of energy flow due to 
exchange of gluons between the subsystems, as well as those with
gluons/quarks/antiquarks outside the system. In other words, in the
spirit of Bak and Sneppen\cite{r16} we are neglecting the dynamical
details {\it totally}.
Having in mind that, 
in such systems,
the gluons as well as the
subsystems ($s_i$'s) of gluons  are {\it virtual}
(space-like), we can ask: 
``How long can such a colorless subsystem
$s_i$ of interacting soft gluons exist,
which carries energy $\varepsilon_i$?''
According to the uncertainty principle, 
the answer should be:
``The time interval
in which the subsystem $s_i$ can exist
is proportional to $1/\varepsilon_i$,
and this quantity can be considered as the lifetime $\tau_i$ of
$s_i$.'' In this sense, those colorless subsystems of gluons are
expected to have larger probabilities to mutate when they are
associated with higher energies, and thus shorter ``lifetimes''.
Note that the basic local interaction
in this self-organized evolution
process is the emission (or absorption) of gluons by gluons prescribed
by the QCD-Lagrangian --- although the detailed mechanisms 
(which can in principle be explicitly written down by
using the QCD-Lagrangian)
do not play a
significant role. 

In terms of the evolution model\cite{r16,r17}
we now call $s_i$ the ``species'' and identify 
the corresponding
lifetime $\tau_i$ as the ``fitness of $s_i$''.
Because of the one-to-one correspondence between $\tau_i$ and 
$\varepsilon_i$, where the latter is a random number,
we can also directly assign random numbers to the $\tau_i$'s
instead. From now we can adopt the evolution model\cite{r16,r17}
and note that,
at the start of such a process (a simulation), the fitness on average
grow, because the least fit are always eliminated. Eventually the
fitness do not grow any further on average. All gluons have a fitness
above some threshold. At the next step, the least fit species (i.e. the
most energetic subsystem $s_i$ of interacting soft gluons), 
which would be right at the threshold, 
will be ``replaced''
and starts an
avalanche (or punctuation of mutation events), which is causally
connected with this triggering ``replacement''. 
After a while, the avalanche will
stop, when all the fitnesses again will be over that threshold. 
In this sense, the evolution goes on, and on, and on.
As in Refs.[\ref{r16}] and [\ref{r17}], we can monitor the duration of
every avalanche, that is the total number of mutation events in
everyone of them, and count how many avalanches of each size are observed.
The
avalanches mentioned here are 
special cases of those discussed in Section \ref{sec:2}.
Their size- and lifetime-distributions are
given by Eq.(\ref{eq1}) and Eq.(\ref{eq2}), 
respectively. Note in particular that the
avalanches in the Bak-Sneppen model correspond to sets of subsystems
$s_i$, the energies ($\epsilon_i$) of which are too high ``to be fit
for the colorless systems of low-energy gluons''. It means, in the
proposed picture, what the virtual photon in deep-inelastic
electron-proton scattering ``meet'' are those ``less fit'' one ---
those who carry ``too much'' energy. 
In a geometrical picture this means, it is 
more probable for such ``relatively energetic'' color-singlet
gluon-clusters ($c_0^\star$) to be spatially
further away from the (confinement region of)
the proton.

There exists, in the mean time, already several versions of evolution 
models\cite{r12,r17} based
on the original idea of Bak and Sneppen\cite{r16}.
Although SOC phenomena have been observed in all these cases\cite{r12,r16,r17},
the slopes of the power-law distributions for the avalanches are different
in different models --- depending on the rules applied to the mutations. 
The values range from  approximately $-1$ to approximately $-2$.
Furthermore, these models\cite{r12,r16,r17} seem to show that neither the
size nor the dimension of the system used for the computer simulation
plays a significant role.

Hence, if we identify 
the colorless charge-neutral object $c_0^\star$ encountered by the
virtual photon $\gamma^\star$ with 
such an avalanche,
we are identifying the
lifetime of $c_0^\star$ with $T$, and the ``size''
(that is the total amount of dissipative energy in this
``avalanche'') with the total amount of energy of $c_0^\star$.
Note that the latter is nothing else but $x_PP^0$, where $P^0$
is the total energy of the proton. This is how and why the
$S$-distribution in Eq. (\ref{eq1}) and the $x_P$-distribution of
$D_S(x_P|\beta,Q^2)$  in Eq.(\ref{eq8}) are related to each other.

\section{The second fingerprint: Temporal scaling}
\label{sec:6}

In this section we discuss in more detail the effects associated with 
the time-degree-of-freedom. In this connection,
some of the concepts and methods related to
the two questions
raised in Section \ref{sec:3} are of great interest. In particular,
one may
wish to know
{\em why} the parton-picture 
does  not work equally well for hadrons and for $c_0^\star$'s.
The answer is very simple: 
The time-degree
of freedom cannot be ignored when we apply
impulse-approximation,
and the applicability of the latter
is the basis of the parton-model.
We recall that,
when we apply the parton model to stable hadrons,
the quarks, antiquarks and
gluons are considered as free and stable objects, 
while the virtual photon $\gamma^\star$ is associated
with a given interaction-time $\tau_{\rm int}(Q^2,x_B)$ characterized
by the values $Q^2$ and $x_B$ of such scattering processes. 
We note however that,  this is possible only when the interaction-time
$\tau_{\rm int}$ is much shorter than the 
corresponding 
time-scales related to hadron-structure
(in particular the average
propagation-time of color-interactions in hadrons).
Having these in mind, we see that, we are confronted with
the following questions when we  deal
with $c_0^\star$'s which have finite lifetimes:
Can we consider the $c_0^\star$'s as ``{\it free}'' and 
``{\it stable}'' particles when
their lifetimes are {\it shorter} than the interaction-time $\tau_{\rm
int}(Q^2,x_B)$? Can we say that a $\gamma^\star$-$c_0^\star$
collision process takes place,
in which the incident $\gamma^\star$ is
absorbed by one or a system of the charged constituents of $c_0^\star$, when
the lifetime $T$ of $c_0^\star$ is {\it shorter} than 
$\tau_{\rm int}(Q^2,x_B)$?

Since the notion
``stable objects'' or ``unstable objects'' depends on the 
scale which is used in
the measurement, the question whether a $c_0^\star$ can 
be considered as a parton
(in the sense that it can be considered as a free 
``stable object'' during the $\gamma^\star$-$c_0^\star$ interaction)
depends very much on the interaction-time
$\tau_{int}(Q^2, x_B)$. 
Here, for
given values of $Q^2$, $x_B$, and thus $\tau_{int}(Q^2, x_B)$,
only those $c^\star_0$'s whose lifetimes ($T$'s) are greater
than $\tau_{int}(Q^2, x_B)$ can absorb the corresponding
$\gamma^\star$.
That
is to say, when we consider diffractive electron-proton scattering in
kinematical regions in which $c_0^\star$'s dominate, 
we must keep in mind that the measured cross-sections (and thus the 
diffractive structure function $F_2^{D(3)}$)
only include contributions from collision-events in which the
condition $T>\tau_{\rm int}(Q^2,x_B)$ 
is satisfied\,! 

\begin{figure}
\psfig{figure=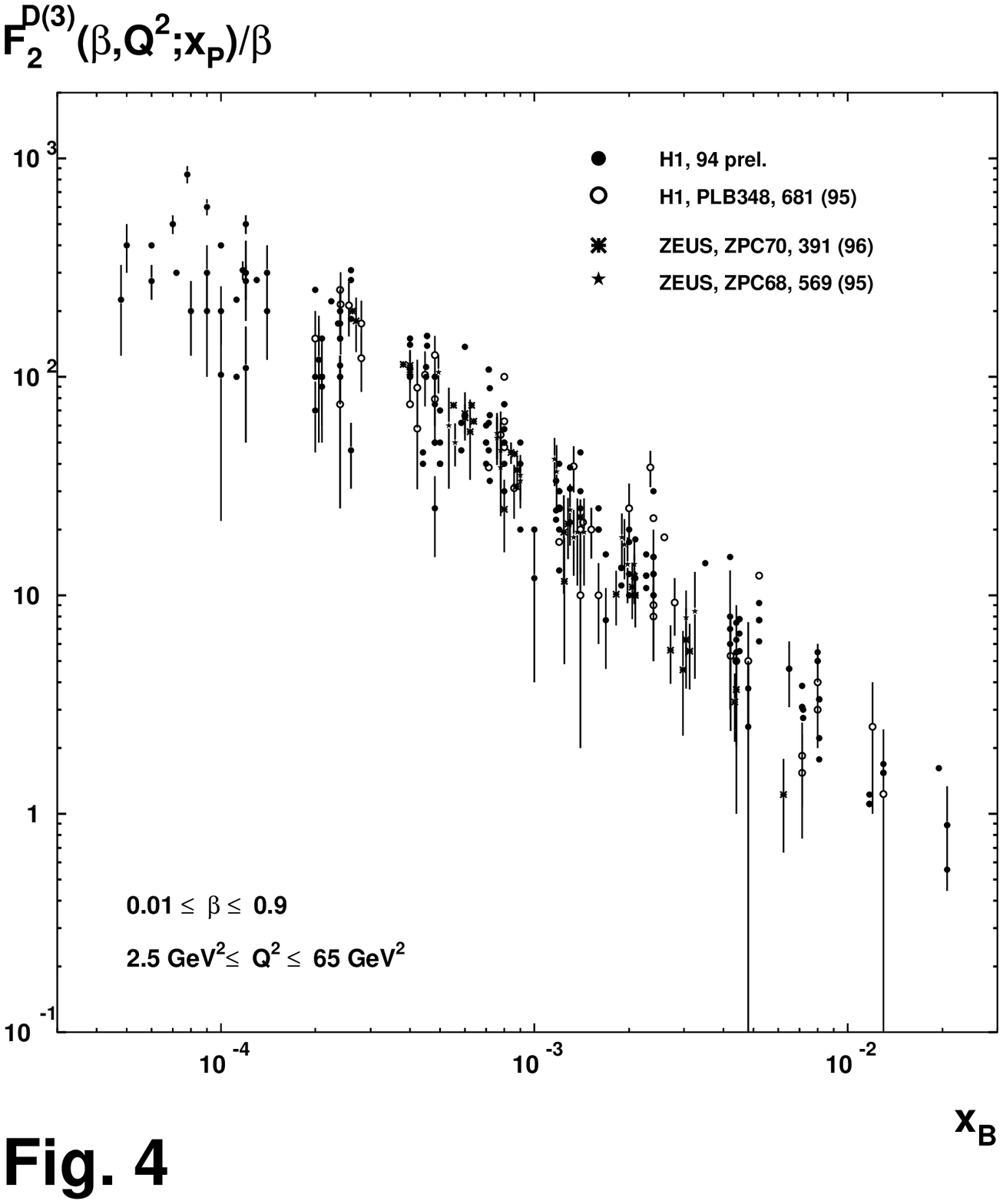,width=16.0cm}
\caption{\label{figure4}
$F_2^{D(3)}(\beta,Q^2;x_P)/\beta$ is plotted as a function of 
$x_B$ in the indicated $\beta$- and $Q^2$-ranges. 
The data are taken from Ref.[\ref{r3}].}
\end{figure}

We note that $\tau_{\rm int}$ can be estimated by making use of the 
uncertainty principle. In fact, by  calculating $1/q^0$ in the 
above-mentioned 
reference frame, 
we obtain
 \begin{equation}
\label{eq10}
\tau_{\rm int}={4|\vec P|\over Q^2}{x_B\over 1-x_B},
\end{equation}
which implies that, for given $|\vec P|$ and $Q^2$ values, 
\begin{equation}
\label{eq11}
\tau_{\rm int}\propto x_B,\hskip 1cm \mbox{\rm for } x_B\ll 1.
\end{equation} 
This means, for diffractive $e^-p$ scattering events in the
small-$x_B$ region at given $|\vec
P|$ and $Q^2$ values, $x_B$ is directly proportional to the interaction
time $\tau_{\rm int}$. Taken together with the relationship between
$\tau_{\rm int}$ and the minimum lifetime $T({\rm min})$ of the
$c_0^\star$'s mentioned above, we reach the following conclusion: The
distribution of this minimum value,
$T({\rm min})$ of the $c_0^\star$'s which dominate the
small-$x_B$ ($x_B<10^{-2}$, say) region can be obtained by examining
the $x_B$-dependence of 
$F_2^{D(3)}(\beta,Q^2;x_P)/\beta$ discussed in
Eqs.(\ref{eq5}), (\ref{eq6}) 
and in Fig.\ref{figure2}. This is because, due to the fact that 
this function is proportional to
the quark (antiquark) distributions $q^c_i(\bar{q_i}^c)$ which can be 
directly probed by the incident virtual photon
$\gamma^\star$, by measuring $F_2^{D(3)}(\beta,Q^2,x_P)/\beta$
as a function of $x_B\equiv \beta x_P$, we are in fact asking
the following question:
Do the distributions of the charged constituents of
$c_0^\star$ depend on the interaction time $\tau_{\rm int}$,
and thus on the minimum lifetime  $T({\rm min})$ of the 
to be detected $c_0^\star$?
We use the identity $x_B\equiv\beta x_P$ and plot the quantity
$F_2^{D(3)}(\beta,Q^2;x_P)/\beta$ against the variable $x_B$
for fixed values of $\beta$ and $Q^2$.
The result of such a log-log plot is given in Fig.\ref{figure4}. It shows
not only  how the dependence on the
time-degree-of-freedom can be extracted from the existing
data\cite{r3}, but also that, for all the measured  
values of $\beta$ and $Q^2$, the quantity
\begin{equation}
\label{eq12}
 p(x_B|\beta, Q^2) \equiv 
{F_2^{D(3)}(\beta, Q^2; x_B/\beta)
\over \beta }
\end{equation}
is approximately
independent of $\beta$, and independent of $Q^2$.
This strongly suggests that the quantity given in Eq.(\ref{eq12})
is associated with some {\em global} features of $c_0^\star$ ---
consistent with the observation made in Section \ref{sec:3} which shows
that it {\em cannot} be used to describe the {\em structure} of $c_0^\star$.
This piece of empirical fact can be expressed by setting
$p(x_B|\beta, Q^2)\approx p(x_B)$. 
By taking a closer look at this $\log$-$\log$ plot, as well
as the corresponding plots for different sets of
fixed $\beta$- and $Q^2$-values (such plots are not
shown here, they are similar to those in Fig.\ref{figure3}),
we see that they are straight lines indicating that 
$p(x_B)$ obeys a power-law. What does this piece of 
experimental fact tell us? What can we learn from
the distribution of the lower limit of the lifetimes (of the
gluon-systems $c_0^\star$'s)? 

\begin{figure}
\psfig{figure=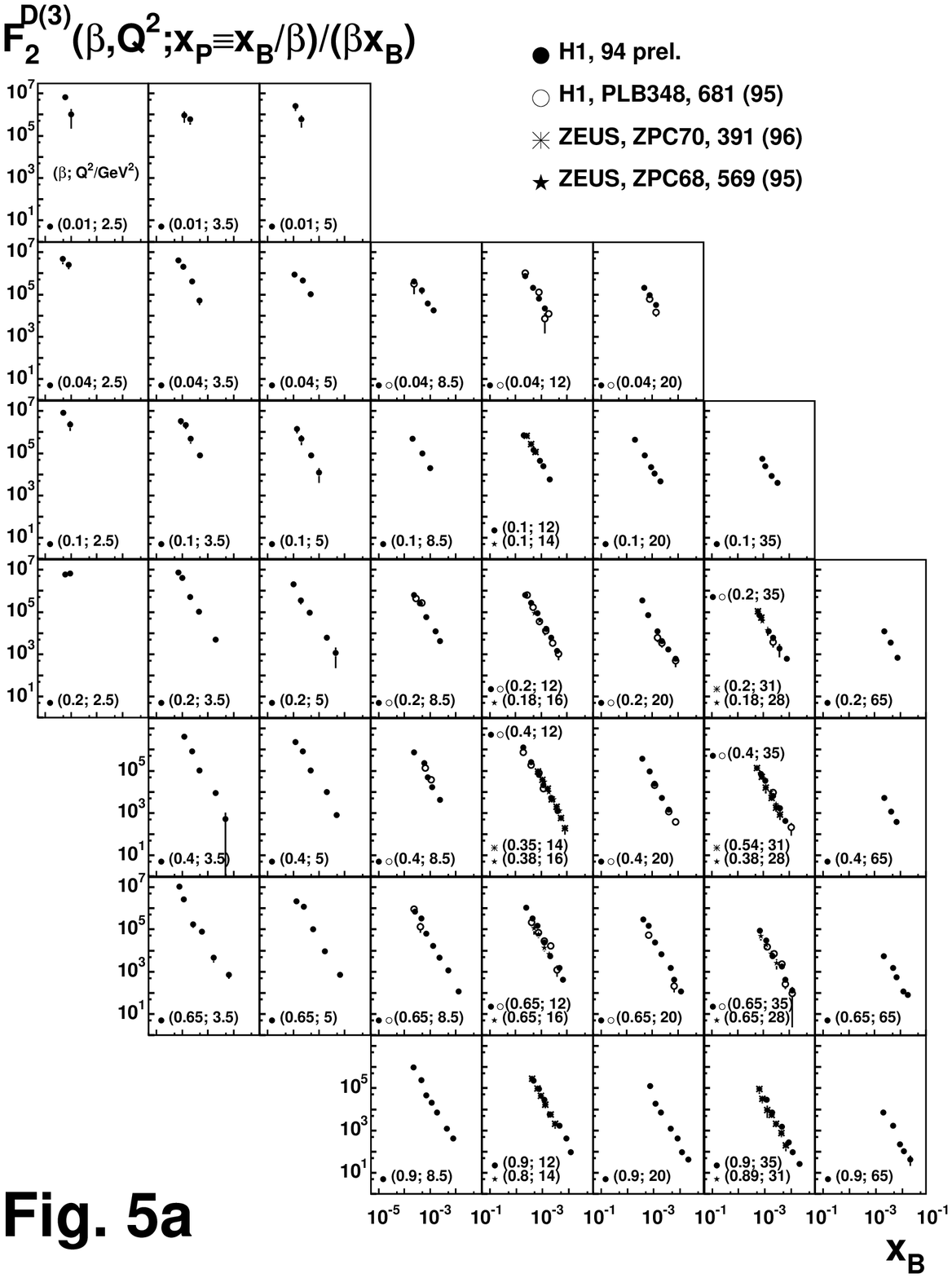,width=16.0cm}
\psfig{figure=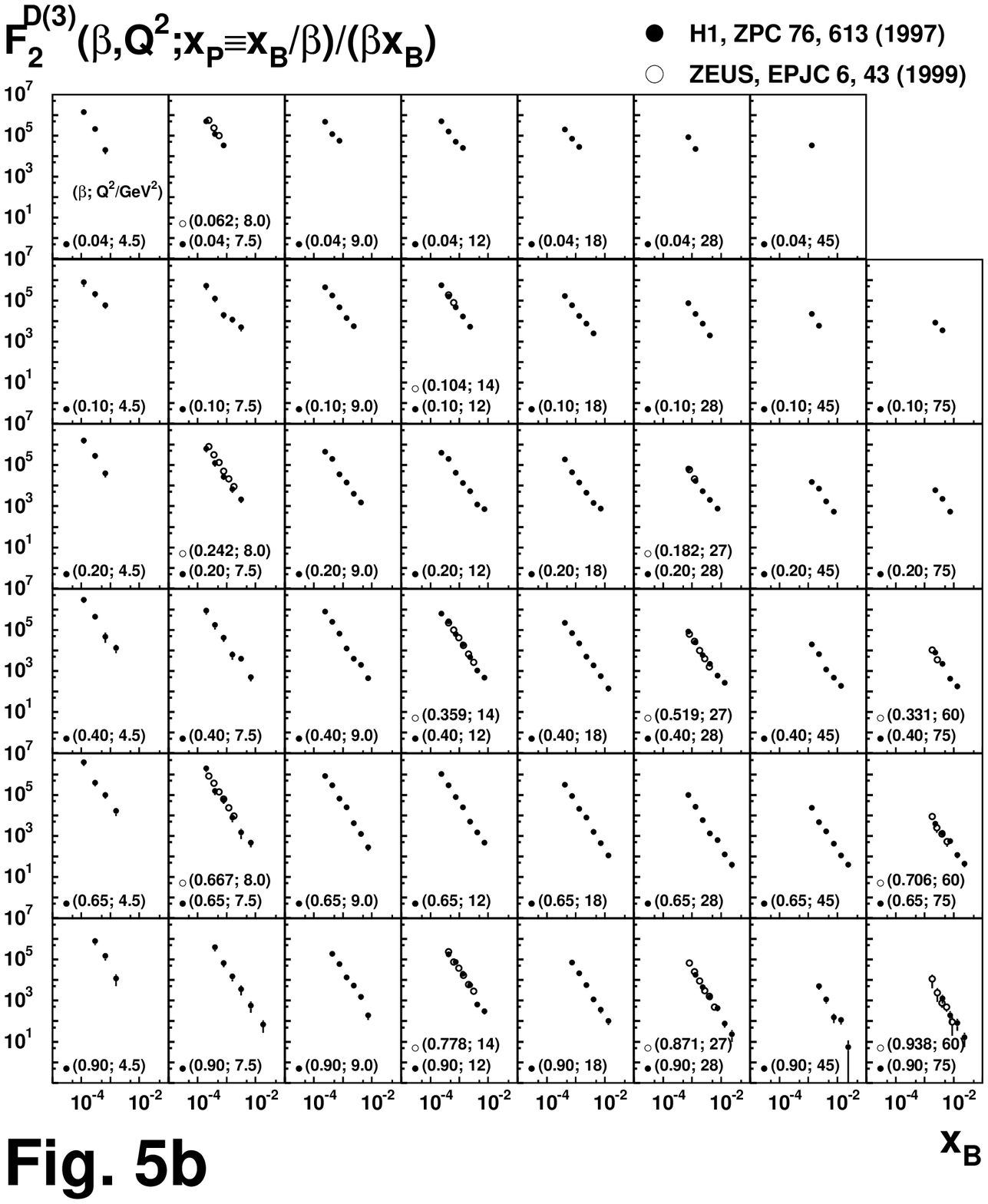,width=16.0cm}
\caption{\label{figure5} 
(a) $F_2^{D(3)}(\beta,Q^2;x_B/\beta)
/(\beta x_B)$ is plotted as a function of 
$x_B$ for fixed  $\beta$- and $Q^2$-values. 
The data are taken from Ref.[\ref{r3}]. 
(b) same as (a) but with data taken from Ref.[\ref{r3a}]. 
Note that in these log-log plots, almost all existing
data points lie on straight-lines with approximately the {\em same}
slope irrespective of the values of $Q^2$ and/or $\beta$. }
\end{figure}

In order to answer these questions, let us,
for a moment, assume that we know the lifetime-distribution $D_T(T)$
of the $c_0^\star$'s. In such a case,
 we can readily evaluate the  integral
\begin{equation}
\label{eq13}
I[\tau_{\rm int}(x_B)]\equiv\int^\infty_{\tau_{\rm int}(x_B)}D_T(T)dT,
\end{equation}
and thus obtain the number density of all those clusters which live 
longer than the interaction time $\tau_{\rm int}(x_B)$.
Hence, under the statistical assumption 
that the chance for a $\gamma^\star$
to be absorbed by one of those
$c_0^\star$'s of lifetime $T$
is proportional to $D_T(T)$ (provided that
$\tau_{\rm int}(Q^2,x_B)\le T$, otherwise this chance is 
zero), we 
can then interpret the integral in Eq.(\ref{eq13}) as follows:
$I[\tau_{\rm int}(Q^2,x_B)]\propto p(x_B)$ is
the probability density for $\gamma^\star$ [associated with the
interaction-time $\tau_{\rm int}(x_B)$] 
to be absorbed by $c_0^\star$'s.
Hence,
\begin{equation}
\label{eq14}
D_T(x_B)\propto {d\over dx_B}p(x_B).
\end{equation}
This means in particular, the fact that $p(x_B)$ [introduced in
Eq.(\ref{eq12})] obeys a power-law in $x_B$ implies that
$D_T(T)$ obeys a  power-law in $T$.
Such a {\em behavior is similar} to 
that shown in Eq.(~\ref{eq2}).
In order to see the {\em quality} of this power-law behavior of $D_T$, and
the {\em quality} of its independence of $Q^2$ and $\beta$, we compare the
above-mentioned behavior with the existing data\cite{r3,r3a}. In
Fig.\ref{figure5}, 
we show the log-log plots of $d/dx_B[p(x_B)]$ against $x_B$. 
In doing this plot, we keep the definition of $p(x_B | \beta ,Q^2)$
given in Eq.(\ref{eq12}) and its weak $\beta$- and $Q^2$-dependence in
mind, and we note that $d/dx_B[p(x_B)]$ is
approximately $F_2^{D(3)}(\beta, Q^2; x_B/\beta)/(\beta x_B)$,
provided that $p(x_B | \beta ,Q^2)$ shows a power-law behavior in
$x_B$. Here, we not only see that the quality of the power-law 
behavior of $D_T$ in $T$ is intimately related to the quality of the
power-law behavior of $F_2^{D(3)}(\beta, Q^2; x_B/\beta)/(\beta x_B)$
in $x_B$, but also how weak the $\beta$- and the $Q^2$-dependence are.

\section{SOC-fingerprints in inelastic diffractive $\gamma^\star p -, 
\gamma p-, pp-,$ and $\bar{p}p-$scattering \\ processes}
\label{sec:7}
We have seen, in Sections \ref{sec:5} and \ref{sec:6}, that in 
diffractive deep-inelastic
electron-proton scattering, the size- and the 
lifetime-distributions of the color-singlet gluon-clusters
($c_0^\star$) obey power-laws,
and that the exponents depend very little
on the variables $\beta$ and $Q^2$. We interpreted
the power-law behaviors as the fingerprints of SOC which are expected
to manifest themselves in systems of interacting soft gluons (which
play the dominating role in diffractive DIS). This expectation is based 
on the fact that the fundamental properties of gluons (in particular
the direct gluon-gluon coupling, the confinement, and the
non-conservation of gluon-numbers) show that the necessary conditions
for the existence of SOC in systems of interacting gluons are
fulfilled, and the fact that (as we can see in various open dynamical
complex systems) power-law behavior in size- and
lifetime-distributions are indeed reliable indicators for the
existence of SOC. In this sense, the existence of such power-law
behavior can be understood in terms of the QCD-based SOC-picture,
although it is not (at least not yet) possible to derive the power-law 
behavior of the size- and lifetime-distributions, and to calculate the
exponents by using non-perturbative QCD. 
But can the observed approximate independence (or weak
dependence) of the exponents on $Q^2$ and $\beta$ {\em also}
be understood in terms of the QCD-based SOC-picture mentioned above? In
particular, what do we expect to see in photoproduction 
processes where the associated value for $Q^2$ is approximately zero?
We note that the possible relationship between the $Q^2>$ a few
GeV$^2$ case and the $Q^2\approx 0$ case in diffractive scattering is
of considerable interest for many reasons. One of them is the fact
that, by comparing these two cases, we can see the fundamental
difference between the conventional (pQCD-corrected parton model plus
Regge phenomenology) picture and the proposed QCD-based SOC-picture
for inelastic diffractive scattering. In the conventional picture, the 
$Q^2>$ a few GeV$^2$ case is ``hard'' and thus should be described by
concepts and methods of parton model and pQCD, while the 
$Q^2\approx 0$ case is ``soft'' and thus should be understood in terms 
of Regge poles. What are the predictions of the proposed SOC-picture? What 
do the experimental data tell us in this connection? 
Would a systematic comparison of the existing data at different
$Q^2$-values --- including those near $Q^2=0$ be useful in
understanding the underlying reaction mechanism(s)
of diffractive scattering in general, and differentiate between the
conventional and the proposed picture in particular?

In order to answer these questions, let us 
recall the space-time aspects of the
collision processes which are closely related
to the above-mentioned
power-law behaviors.
Viewed in a fast moving frame (e.g. the c.m.s. of the colliding
electron and proton), the states of the interacting soft gluons 
originating from  the
proton are self-organized.
The $c_0^\star$ caused by local perturbations
and developed through ``domino effects'' are BTW-avalanches 
(see Sections \ref{sec:1} and \ref{sec:5}), the 
size-distribution of which [see Eqs.(\ref{eq8}) and (\ref{eq1})] are given by
Fig.\ref{figure3}. This explicitly shows that
there are $c_0^\star$'s of all sizes,
because a power-law
size-distribution implies that there is no scale in size.
Recall that, since such $c_0^\star$'s are color-singlets, their 
spatial extensions can be much larger than that of the proton,
and thus they can  be ``seen'' also {\em outside} the proton 
by a virtual
photon originating from the electron. 
In other words, what the virtual photon encounters is a cloud of 
$c_0^\star$'s, everyone of which is in general partly inside 
and partly outside the proton.

The virtual photon, when it encounters a $c_0^\star$, will be absorbed 
by the charged constituents
(quarks and antiquarks due to fluctuations of the gluons)
of the gluon-system. Here it is useful to recall that in such a space-time
picture, $Q^2$ is inversely proportional to the transverse size,
and $x_B$ is a measure of the interaction time 
[See Eqs. (\ref{eq10}) and (\ref{eq11})
in Section \ref{sec:6}] of the virtual photon.
It is conceivable, that the values for the cross-sections for virtual
photons (associated with a given $Q^2$ and a given $x_B$) to 
collide with $c_0^\star$'s (of a given size and a given 
lifetime) may depend on these variables. But, since the
processes of self-organization (which produce such $c_0^\star$'s) 
take
place independent of the virtual photon (which originates from the 
incident electron and enters ``the cloud'' to
look for suitable partners), the power-law behaviors of the size-
and lifetime-distributions of the $c_0^\star$'s are expected to be
independent of the properties associated with the virtual photon.
This means, by using
$\gamma^\star$'s associated with different values 
of $Q^2$ to detect $c_0^\star$'s of various sizes, 
we are moving up or down on the straight lines in the 
log-log plots for the size- and lifetime distributions, 
the slopes of which do not change.
In other words, the observed
approximative $Q^2$-independence of the slope in the above-mentioned
log-log plots of the data can be considered as 
a natural consequence of the QCD-based SOC picture.

As far as the $\beta$-dependence is concerned, we recall the
results obtained  in Sections \ref{sec:3} and \ref{sec:4},
which explicitly show the following:
The $c_0^\star$'s
{\em cannot} be considered as {\em hadrons}. In particular, it is
neither possible
nor meaningful
to talk about ``{\em the} electromagnetic structure of {\em the}
(or {\em a typical}, or {\em an average}) $c_0^\star$''. 
This is not only because the power-law behavior of the size and the
lifetime distributions of such $c_0^\star$'s implies that such
objects --- although they are color-singlets --- can have neither a
typical (an average) size, nor a typical (an average) lifetime, but
also because of the following fact: When $F_2^{D(3)}(\beta , Q^2;
x_P )/\beta$ which is usually known \cite{r8} as ``the
electromagnetic structure function of the colorless object exchanged
in diffractive scattering processes'' is plotted as functions of
$\beta$ (cf. Fig.\ref{figure2}), we see a rather significant
$x_P$-dependence. This {\em does not mean}, however, that the measured 
$\beta$-dependence of $F_2^{D(3)}$ {\em cannot} provide us with {\em any}
further information on the electromagnetic properties of the
color-singlet gluon-clusters ($c_0^\star$)! 
This is because, the $c_0^\star$'s
which play the dominating role in diffractive scattering are
color-singlets, hence, even when such clusters are BTW-avalanches
which have neither a typical size nor a typical lifetime, a set of
such clusters with given size- and lifetime-distributions can
nevertheless be considered as {\em a specific set of hadrons with distinct
properties}. Hence, when the $\beta$-dependence of  $F_2^{D(3)}(\beta , Q^2;
x_P )/\beta$ is examined 
in inelastic diffractive scattering processes, the
electromagnetic properties of such a set of hadrons are probed by the
incident (virtual or real, depending on the $Q^2$-value of the event)
photons. In this connection, it is perhaps useful to consider the
$\beta$-distribution integrated over $x_P$

\begin{figure}
\psfig{figure=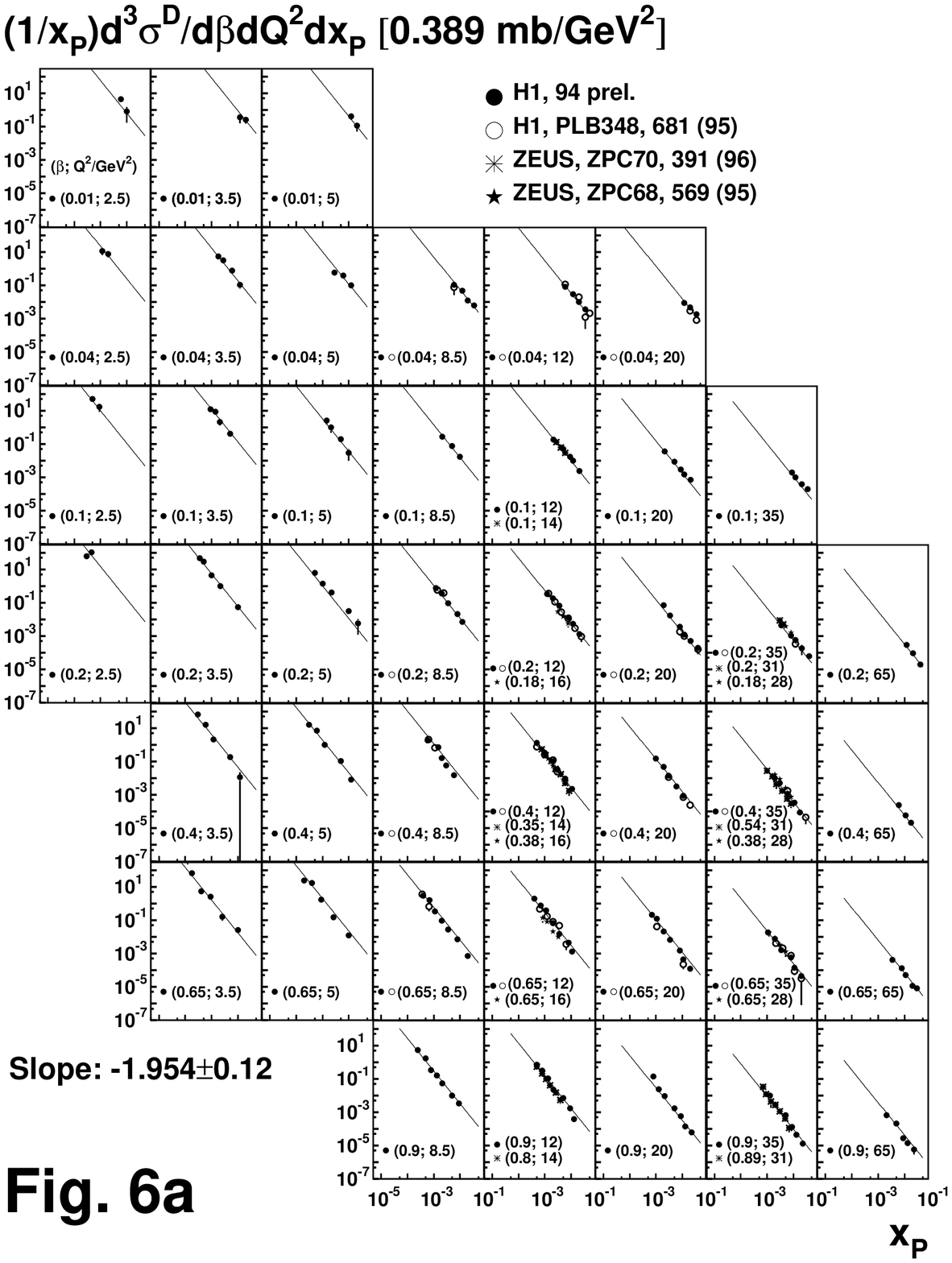,width=16.0cm}
\psfig{figure=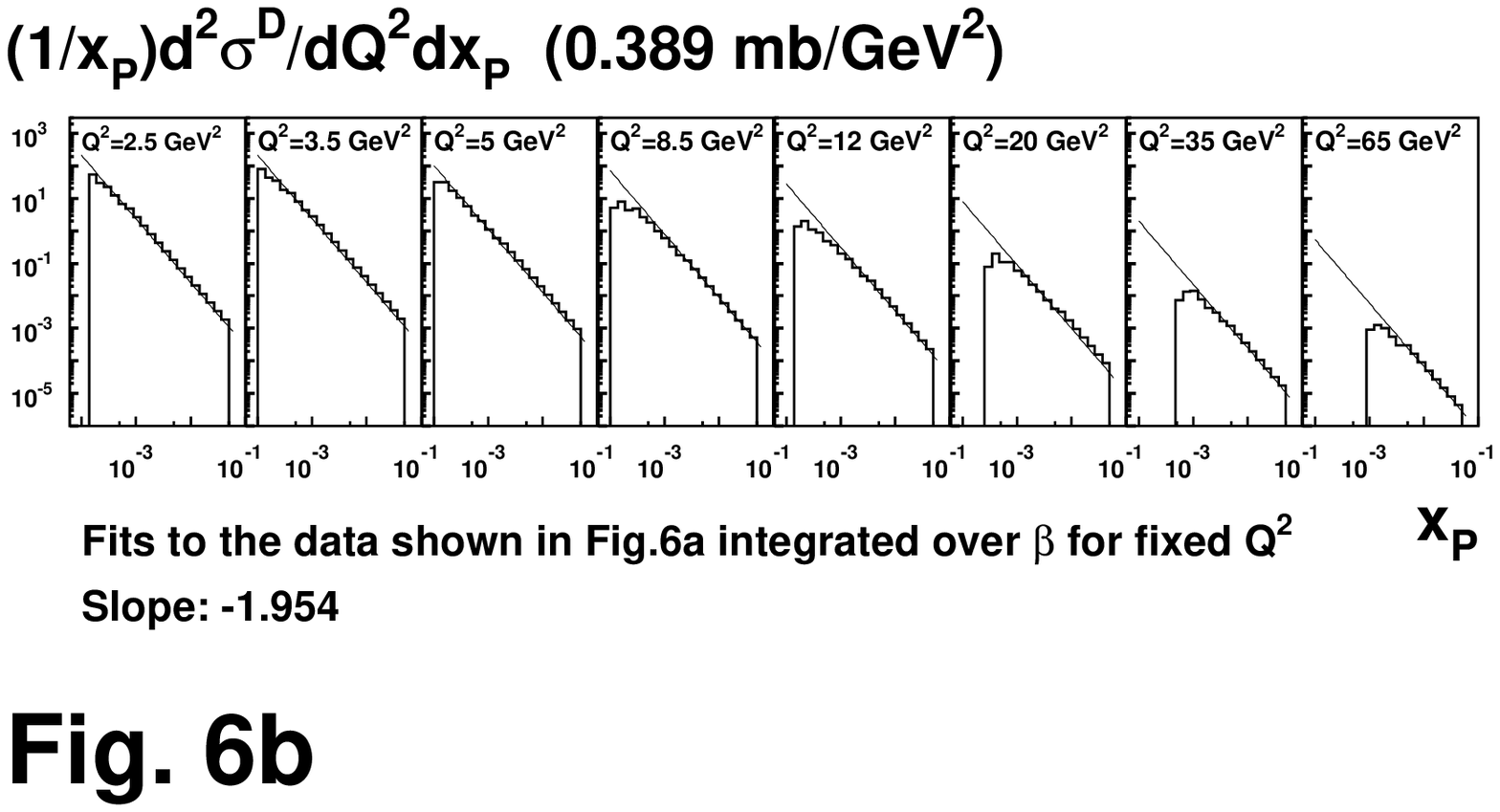,width=16.0cm}
\psfig{figure=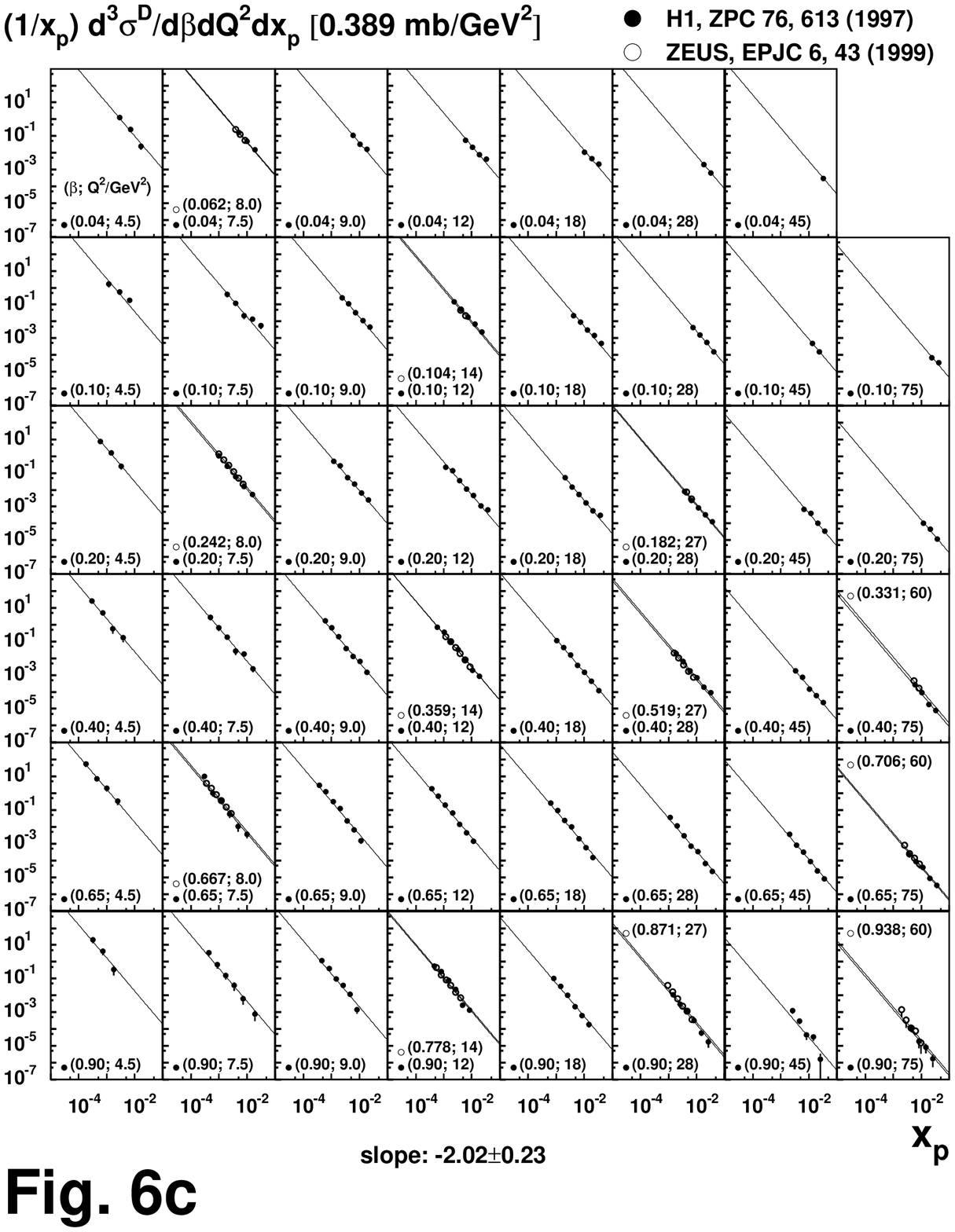,width=16.0cm}
\caption{\label{figure6}
            (a) $(1/x_P)d^3\sigma^D/d\beta dQ^2 dx_P$ 
            is plotted as a function of $x_P$ in different
            bins of $\beta$ and $Q^2$. The data are taken from
            Ref.[\ref{r3}]. (The factor $0.389\,\mbox{mb}$ is due to
            $(1\,\mbox{GeV})^{-2} = 0.389\,\mbox{mb}$.)
         (b) $(1/x_P)d^2\sigma^D/ dQ^2 dx_P$ in units of
             $\mbox{GeV}^{-4}$ 
            is plotted as a function of $x_P$ in different
            bins of $Q^2$. The data are taken from
            Ref.[\ref{r3}].
         (c) same as (a) but with data taken from Ref.[\ref{r3a}].
             (The factor $0.389\,\mbox{mb}$ is due to
            $(1\,\mbox{GeV})^{-2} = 0.389\,\mbox{mb}$.)}
\end{figure}

For the purpose of comparing SOC-fingerprints obtained at different
$Q^2$-values, we are interested much more in 
measurable quantities in which the integrations over $\beta$ 
have been carried out.
A suitable candidate for this purpose is the differential cross-section
\begin{eqnarray}
\label{eq15}
\lefteqn{\frac{1}{x_P}\,\frac{d^2\sigma^D}{dQ^2 dx_P} = } \nonumber \\
& = &\int d\beta \,\frac{4\pi\alpha^2}{\beta Q^4}\,
            \left( 1-y+\frac{y^2}{2}\right)\,
            \frac{F_2^{D(3)}(\beta, Q^2; x_P)}{x_P} \nonumber\\
& \approx &
\int d\beta \,\frac{4\pi\alpha^2}{\beta Q^4}\,
            \left( 1-y+\frac{y^2}{2}\right)\,
             D_S(x_P| \beta, Q^2)
\end{eqnarray}
Together with Eqs.(\ref{eq3}) and (\ref{eq8}), 
we see that this cross-section is 
nothing else but the effective $\beta$-weighted
$x_P$-distribution $D_S(x_P|Q^2,\beta)$ of the 
$c_0^\star$'s. Note that the weighting factors shown on the 
right-hand-side of Eq.(\ref{eq15}) are simply results of QED! 
Next, we use the data\cite{r3,r3a} for 
$F_2^{D(3)}$ which are available at present,
to do a log-log plot for the integrand of the expression
in Eq.(\ref{eq15}) as a function of $x_P$
for different values of $\beta$ and $Q^2$.
This is shown in 
in Fig.\ref{figure6}a and Fig.\ref{figure6}c. 
Since the absolute values of this quantity depend
very much, but the slope of the curves very little on $\beta$,
we carry out the integration as follows:
We first fit every set of the data separately.
Having obtained the slopes and the intersection points,
we use the obtained fits to perform the integration over $\beta$.
The results are shown in the
\begin{eqnarray*}
\log{\left(\frac{1}{x_P}\,\frac{d^2\sigma^D}{dQ^2\,dx_P}\right)}
& \mbox{\ \ versus\ \ \ } &
\log{(x_P)}
\end{eqnarray*}
plots of Fig.\ref{figure6}b.
These results show the $Q^2$-dependence of the slopes
is practically negligible, and that the slope
is approximately $-1.95$ for all values of $Q^2$.

\begin{figure}
\psfig{figure=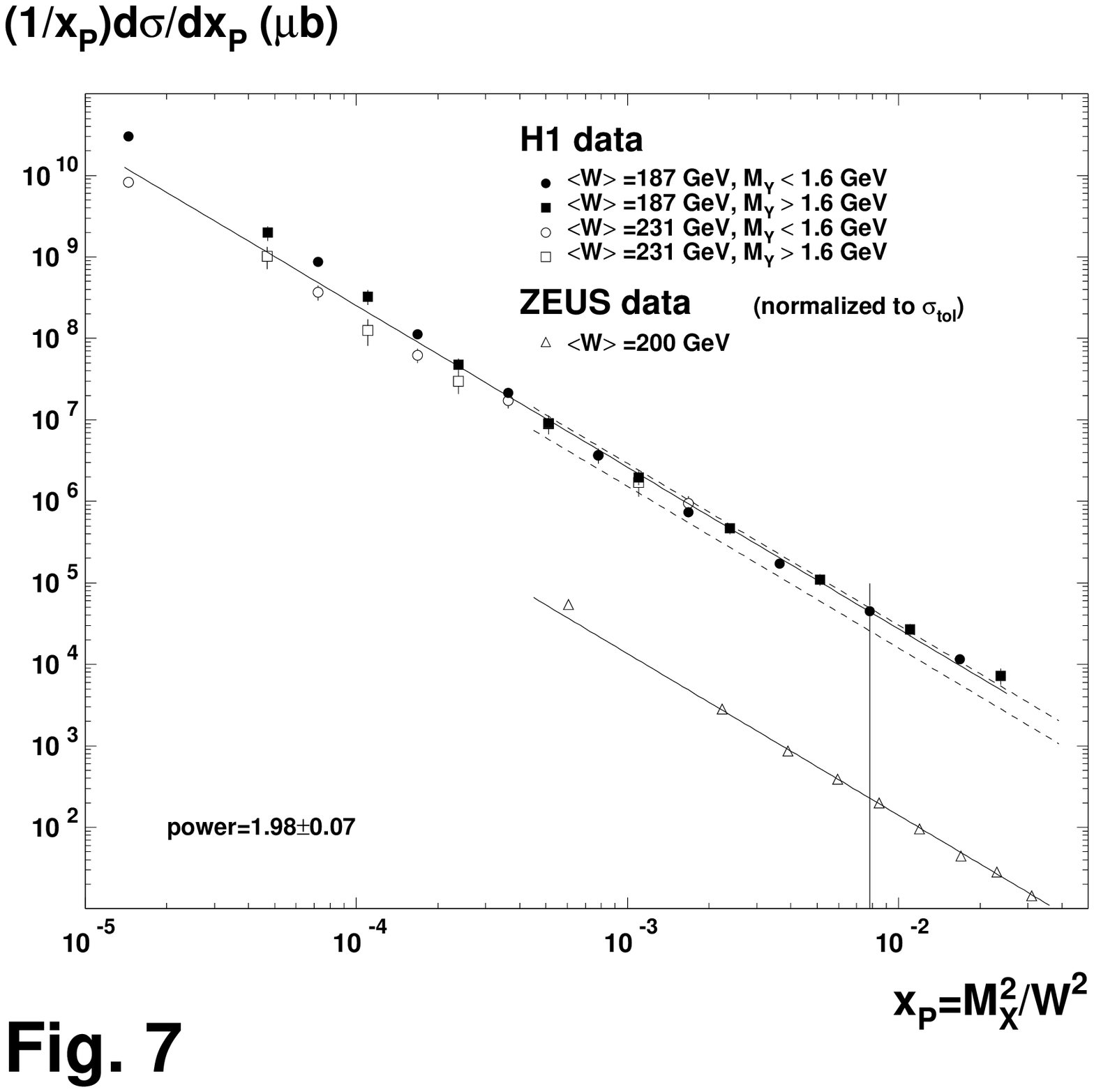,width=16.0cm}
\caption{ \label{figure7}
           $(1/x_P)d\sigma/dx_P$ for photoproduction
          $\gamma + p \rightarrow X + p$ 
            is plotted as a function of $x_P$.
            The data are taken from Ref.[\ref{r18}].
           Note that the data in the second paper are given in terms of 
           relative cross sections. Note also that the slopes of the 
           straight-lines are the same. The two dashed lines indicate
           the lower and the upper limits of
           the results obtained by multiplying the lower solid line
           by $\sigma_{\mbox{tot}}=154\pm 16 \mbox{(stat.)} \pm
            32 \mbox{(syst.)} \mu b$. This value is taken from the third
           paper in Ref.[\ref{r18}].}
\end{figure}

Furthermore, in order to see whether the quantity introduced in
Eq.(\ref{eq15}) is indeed
useful, and in order to perform a decisive test of the 
$Q^2$-independence of the slope in the power-law behavior
of the above-mentioned size-distributions,
we now
compare the results in deep-inelastic
scattering\cite{r3,r3a} with those obtained in photoproduction\cite{r18},
where LRG events have
also been observed. This means, as in
diffractive deep-inelastic scattering, we again associate the
observed effects with colorless objects which are interpreted as
a system of interacting soft gluons originating from the proton.
In order to find out whether it is the same kind of 
gluon-clusters as in deep-inelastic scattering, and whether
they ``look'' very much different when we probe them with
real ($Q^2=0$) photons, we replot the existing
$d\sigma/dM_x^2$ data\cite{r18} for photoproduction 
experiments performed at different total energies,
and note
the kinematical relationship between $M_x^2$, $W^2$ and $x_P$
(for $Q^2\ll M^2$ and $|t|\ll M_x^2$):
\begin{eqnarray}
\label{eq16}
x_P \approx \frac{M_x^2}{W^2} & &
\end{eqnarray}
The result of the corresponding
\begin{eqnarray*}
\log{\left(\frac{1}{x_P}\,\frac{d\sigma}{dx_P}\right)}
& \mbox{\ \ versus\ \ \ } &
\log{(x_P)}
\end{eqnarray*}
plot is shown in Fig.\ref{figure7}. The slope obtained from  a least-square
fit to the existing data\cite{r18} is $-1.98\pm 0.07$.

The results obtained in diffractive
deep-inelastic electron-proton scattering
and that for diffractive photoproduction strongly suggest
the following: The formation processes of $c_0^\star$'s
in the proton is due to self-organized criticality, and thus
the spatial distributions of such clusters
--- represented by the $x_P$-distribution ---
obey power-laws.
The exponents of 
such power-laws are
independent of
$Q^2$. Since $1/Q^2$ can be interpreted
as a measure for the transverse 
size  of the incident virtual photon, the observed 
$Q^2$-independence of the exponents can be
considered as further evidence for SOC ---
in the sense that the self-organized gluon-cluster formation
processes take place independent of the 
photon which is ``sent in'' to detect the clusters.

Having these results, and the close relationship between real photons
and hadrons in mind, we are immediately led to the following questions:
What shall we see, when we replace the (virtual or real) photon by a
hadron --- a proton or an antiproton? (See in this connection
Fig.\ref{figure8}, 
for the notations and the kinematical relations for the description of 
such scattering processes.) Should we not see similar behaviors, if
SOC in gluon-systems is indeed the reason for the occurrence of
$c_0^\star$'s which can be probed experimentally in
inelastic diffractive scattering processes? To answer these questions,
we took a closer look at the available single diffractive
proton-proton and proton-antiproton scattering data\cite{r4,r5};
and in
order to make quantitative comparisons, we plot the quantities
which correspond to those shown in Fig.\ref{figure6}b and Fig.\ref{figure7}.
These plots are shown
in Fig.\ref{figure9}a and Fig.\ref{figure9}b.
In Fig.\ref{figure9}a, we see the double differential
cross-section $(1/x_P)d^2\sigma/(dtdx_P)$ at four different
$t$-values.
In Fig.\ref{figure9}b, 
we see the integrated differential cross-section
$(1/x_P)d\sigma/dx_P$.
Note that, here 
\begin{equation}
\label{eq17}
x_P \approx M_x^2/s,
\end{equation}
where $\sqrt{s}$ is the total
c.m.s. energy of the colliding proton-proton or antiproton-proton system.
Here, the integrations of
the double
differential cross-section 
over $t$ are in the ranges in which 
the corresponding experiments have been performed.
(The extremely
weak energy-dependence has been ignored in the integration.)
The
dashed lines in all the plots in Figs.\ref{figure9}a and
\ref{figure9}b 
stand for the slope
$-1.97$ which is the average of the slope obtained from the plots
shown in Figs.\ref{figure6}b and \ref{figure7}.
This means, the result shows exactly what we expect to see: The
fingerprints of SOC can be clearly seen 
also in proton- and antiproton-induced
inelastic diffractive scattering processes, 
showing that such characteristic features 
are indeed universal and robust!

We are thus led to the following conclusions. Color-singlet
gluon-clusters ($c_0^\star$) can be formed in hadrons as a consequence of
self-organized criticality (SOC) in systems of interacting soft
gluons. In other words, ``the colorless objects'' which dominate the
inelastic diffractive scattering processes are BTW-avalanches
(BTW-clusters). Such $c_0^\star$'s are in general distributed
partly inside and partly outside the confinement region of the
``mother-hadron''. Since the interactions between the $c_0^\star$'s
and other color-singlet objects (including the target
proton) should be of Van der Waals type, it is expected that such an
object can be
readily driven out of the above-mentioned confinement region
by the beam-particle in geometrically more peripheral collisions. This
has been checked by examing inelastic single-diffractive
scattering processes at high energies in which virtual photon, real
photon, proton, and antiproton are used as beam particles. The result
of this systematic check shows that the universal distributions of such
$c_0^\star$'s can be directly extracted 
from the data. In particular,
the fact that $x_P$ is the energy fraction carried by the struck
$c_0^\star$'s, and the fact that the $x_P$-distributions are 
universal, it
is tempting to regard such $x_P$-distributions as ``the
parton-distributions''  
for diffractive scattering processes. Having seen this, it is also
tempting to ask: Can we make use of such
``parton-distributions'' to describe and/or to predict the measurable
cross-sections in inelastic diffractive scattering processes? This and 
other related questions will be discussed in the following sections.



\section{Diffractive scattering in high-energy collisions and
diffraction in optics}
\label{sec:8}
It might sound strange, but it is true that people working in this
field of physics often do not agree with one another on the
question:
What is ``Diffractive Scattering in High-Energy Collisions''? In this paper,
we have, until now, simply adopted the currently \cite{r8} 
popular definition of
``inelastic diffractive scattering processes''. That is, 
when we talked about ``inelastic diffractive scattering'' we
were always referring to processes in which
``colorless objects'' are ``exchanged''. 
In other words, until now,
the following question has {\em not} been asked: 
Are the above-mentioned ``inelastic {\em diffractive} scattering
processes'' indeed comparable with {\em diffraction} in
Optics, in the sense that the beam particles should be considered
as waves, and the target-proton together with the associated 
(in whatever manner) colorless objects can indeed be viewed as a 
``scattering screen''? 

This question will be discussed in the present and the subsequent
sections, together with the existing data \cite{r4,ppbardata} for 
the double differential cross-section $d^2\sigma/dt\,d(M_x^2/s)$ for 
proton-proton and antiproton-proton collisions 
(where $t$ is the 
4-momentum-transfer squared, $M_x$ is the missing-mass, and $\sqrt{s}$ is the 
total c.m.s. energy). The purpose of this investigation is to 
find out: ``Can the observed 
$t$-dependence and the $(M_x^2/s)$-dependence of  $d^2\sigma/dt\,d(M_x^2/s)$ 
in the given kinematic range  
($0.2\mbox{\,GeV}^2$$\le$$|t|$$\le$ $3.25 \mbox{\,GeV}^2$,
$16\mbox{\,GeV}$$\le$$\sqrt{s}$$\le$ $630\mbox{\,GeV}$, 
and $M_x^2/s$$\le$ $0.1$)
be understood in terms of the well-known concept of diffraction in Optics? 
The answer to this question is of particular interest for several
reasons:

(a) High-energy proton-proton and proton-antiproton scattering at
small momentum transfer has played, and is still playing a very
special role in understanding diffraction
and/or diffractive dissociation in lepton-, photon-
and hadron-induced 
reactions\cite{r2,r3,r3a,r4,r5,r7,r8,r18,ppbardata,ReviewSmallt,OpticalModels}. 
Many
experiments
have been
performed at various incident energies for elastic and inelastic
diffractive scattering processes. It is known that the double
differential cross section $d^2\sigma/dt\,d(M_x^2/s)$ is a quantity
which can yield much information on the reaction
mechanism(s) and/or on the structure of the participating colliding 
objects. In the past, the 
$t$-, $M_x$- and $s$-dependence of the differential cross-sections for
inelastic diffractive scattering processes
has been presented in different forms,
where a number of interesting features have been 
observed\cite{r4,ppbardata,ReviewSmallt}.
For example, it is seen that, the $t$-dependence of 
$d^2\sigma/dt\,dM_x^2$ at fixed $s$
depends very much on $M_x$; 
the $M_x^2$-dependence of 
$d^2\sigma/dt\,dM_x^2$
at fixed $t$ depends on $s$.
But, when $d^2\sigma/dt\,d(M_x^2/s)$ is plotted as
function of $M_x^2/s$ at 
given $t$ -values (in the range 
$0.2\,\mbox{GeV}^2\le |t|\le 3.25\,\mbox{GeV}^2$)
they are approximately independent of
$s$! What do these observed striking features tell us?
The first precision
measurement of this quantity was published more than twenty years 
ago\cite{r4}. 
Can this, as well as the more recent
$d^2\sigma/dt\,d(M_x^2/s)$-data\cite{ppbardata} be understood theoretically? 

(b) The idea of using optical and/or geometrical analogies to describe 
high-energy hadron-nucleus and hadron-hadron collisions at small
scattering angles has been discussed by many authors
\cite{OpticalModels,ReviewSmallt} many years ago. It is shown in
particular that this approach is very useful and 
successful in describing elastic scattering.
However, it seems that, until now, no attempt has been made 
to describe the  data\cite{r4,ppbardata} by
performing 
quantitative calculations for 
$d^2\sigma/dt\,d(M_x^2/s)$
by using optical geometrical analogies.
It seems worthwhile to make such an attempt. This is because, it
has been pointed out\cite{r10a} very recently, that 
the above-mentioned analogy
can be made to understand the observed $t$-dependence in
$d\sigma /dt$.

(c) Inelastic diffractive $pp$- and $\bar{p}p$-scattering
belongs to those soft processes which have also been extensivly discussed
in the well-known Regge-pole approach\cite{r7,r8,ReviewSmallt}. 
The basic idea of this approach is that colorless objects in form of
Regge trajectories (Pomerons, 
Reggeons etc.) are exchanged during the collision,
and such trajectories are responsible for the dynamics of the
scattering processes.
In this approach, it is
the $t$-dependence of the
Regge trajectories, the $t$-dependence 
of the corresponding
Regge residue functions, the properties of the coupling of the
contributing trajectories (e.g. triple Pomeron or  
Pomeron-Reggeon-Pomeron coupling),
and the number of contributing Regge trajectories which
determine the experimentally observed $t$- and  $M_x$-dependence of
$d^2\sigma/dt\,d(M_x^2/s)$.  
A number of Regge-pole models\cite{r8,r7} 
have been 
proposed, and there exist good fits \cite{r8,r7} to the 
data. What remains to be understood in this approach is the dynamical
origin of the Regge-trajectories on the one hand, and the physical
meaning of the unknown functions (for example the $t$-dependence of
any one of the Regge-residue functions) on the other.
It has been pointed out \cite{r17a,r10a}, that there may 
be an overlap between the ``Partons in Pomeron and Reggeons'' picture
and the SOC-picture \cite{r10a}, and that one way 
to study the possible relationship between the two approaches 
is to take a closer look at the double differential cross-section
$d^2\sigma/dt\,d(M_x^2/s)$.

\section{Optical diffraction off dynamical complex systems}
\label{sec:9}
Let us begin our discussion on the above-mentioned questions by
recalling that the concept of ``diffraction'' or ``diffractive scattering''
has its origin in Optics, and Optics is part of Electrodynamics, 
which is not only the {\em classical limit}, but also {\em the
basis} of Quantum Electrodynamics (QED).
Here, it is useful to recall in particular the following:
Optical diffraction is associated with 
departure from geometrical optics 
caused by the finite wavelength of light.
Frauenhofer diffraction can be observed by placing  
a scatterer (which can in general be a
scattering screen with more than one aperture or a 
system of scattering objects) 
in  the path of propagation of light (the wavelength 
of which is less than the linear dimension of the scatterer) where 
not only
the light-source, but also the detecting device, are very far 
away from the 
scatterer. The parallel incident light-rays can be 
considered as plane waves (characterized by a set of constants 
$\vec k, w\equiv |\vec k|$, and $u$ say, which denote
the wave vector, the frequency and the  amplitude of a component of the  
electromagnetic field respectively in the laboratory frame).  
After the scattering, 
the scattered field can be written in accordance with Huygens' principle as
\begin {equation} 
\label{eq18}
u_P = \frac{e^{i|\vec{k}^\prime|R}}{R} f(\vec{k},\vec{k}^\prime).
\end{equation}
Here, $u_P$ stands for a component of the field originating from 
the scatterer, $\vec{k}^\prime$ is the wave vector of the scattered light 
in the direction of observation, $|\vec{k}^\prime|\equiv \omega^\prime$ 
is the corresponding frequency, $R$ is the distance between the 
scatterer and the observation point $P$, and 
$f(\vec{k},\vec{k}^\prime$) is the (unnormalized) scattering 
amplitude which describes the change of the wave vector in the 
scattering process. By choosing a coordinate system in which the 
$z$-axis coincides with the incident wave vector $\vec{k}$, 
the scattering amplitude can be expressed as 
follows\cite{Landau,OpticalModels,ReviewSmallt}
\begin{eqnarray}
\label{eq19} 
f(\vec q) & = & 
\frac{1}{(2\pi)^2} \int\!\int\limits_{\mbox{\hspace*{-0.5cm}}\Sigma}^{} 
              d^2\vec{b}\,\alpha(\vec{b})\,
              e^{-i\vec q \cdot \vec{b}}\mbox{\ .}
\end{eqnarray}
Here,  $\vec{q}\equiv \vec{k}^\prime - \vec{k}$ determines 
the change in wave vector due to diffraction; $\vec{b}$ is the 
impact parameter which indicates the position of 
an infinitesimal surface element on the wave-front ``immediately
behind the scatterer'' where the incident wave would reach in the
limit of geometrical optics, and $\alpha(\vec{b})$ 
is the corresponding amplitude 
(associated with the boundary conditions which the scattered field
should satisfy)
in the two-dimensional
impact-parameter-space (which is here the $xy$-plane), and the
integration extends over the region $\Sigma$ in which
$\alpha(\vec{b})$ is different from zero. In those cases in which the 
scatterer is symmetric with respect to the scattering axis (here the $z$-axis),
Eq.(\ref{eq19}) 
can be expressed, by using an integral representation 
for $J_0$, as
\begin{eqnarray}
\label{eq20}
f(q) & = & \frac{1}{2\pi} \int\limits_{0}^{\infty} b\, db\, 
                 \alpha(b) J_0(q b)\mbox{\ .}
\end{eqnarray}
where $q$ and $b$ are the magnitudes of $\vec{q}$ and
$\vec{b}$ respectively.

The following should be mentioned in connection with Eqs.(\ref{eq19}) 
and (\ref{eq20}):
Many of the well-known phenomena related to Frauenhofer diffraction
have been deduced\cite{Landau} from these equations under the additional
condition (which is directly related to the boundary conditions
imposed on the scattered field)
$|\vec{k}^\prime|=|\vec{k}|=\omega^\prime=\omega$, 
that is,  $\vec{k}^\prime$ differs 
from $\vec{k}$ only in direction.
In other words, the outgoing light wave has exactly the same frequency, 
and exactly the same magnitude of wave-vector
as those for the 
incoming wave. (This means, quantum mechanically speaking, 
the outgoing photons are also on-shell photons, the
energies of which are the same as the incoming ones.)
In such cases, it is possible to envisage that
$\vec{q}$ is approximately perpendicular to $\vec{k}$ and to 
$\vec{k}^\prime$, that is, $\vec{q}$
is approximately
perpendicular to the chosen $z$-axis, and thus in the above-mentioned
$xy$-plane (that is $\vec{q}\approx\vec{q}_\perp$).
While the scattering angle distribution in such 
processes (which are considered as the characteristic features of
{\em elastic}
diffractive scattering) 
plays a significant role in understanding 
the observed diffraction phenomena, it
is of considerable importance to note the following:

(A) Eqs.(\ref{eq19}) and (\ref{eq20}) can be used 
to describe diffractive scattering
with, or without, this additional
condition, provided that the difference of 
$\vec{k}^\prime$ and $\vec{k}$ in the 
longitudinal direction (i.e. in the direction
of $\vec{k}$) is small compared to 
$q_{\perp}\equiv |\vec{q}_{\perp}|$ 
so that 
$\vec{q}_\perp$ can be approximated by $\vec{q}$. In fact,
Eqs.(\ref{eq19}) and 
(\ref{eq20}) are strictly valid when $\vec{q}$ is a vector in the
above-mentioned $xy$-plane, that is when we write $\vec{q}_\perp$
instead of $\vec{q}$. 
Now, since Eqs.(\ref{eq19}) and (\ref{eq20}) in such a form 
(that is when the replacement $\vec{q}\rightarrow \vec{q}_\perp$ is made)
are valid {\em without} the condition $\vec{q}$ should approximately
be equal to $\vec{q}_\perp$ and in particular without the additional condition 
$|\vec{k}^\prime|=|\vec{k}|=\omega^\prime=\omega$,
it is clear that they are also valid for {\em inelastic}
scattering processes.
In other words, Eqs.(\ref{eq19}) and (\ref{eq20}) 
can also be used to describe {\em inelastic}
diffractive scattering (that is, processes
in which 
$\omega^\prime\neq\omega$,
$|\vec{k}^\prime|\neq|\vec{k}|$)
provided that the following replacements are made.
In Eq.(\ref{eq19}), $\vec{q}\rightarrow \vec{q}_\perp$,
$f(\vec{q})\rightarrow 
f_{\mbox{\tiny inel.}}(\vec{q}_\perp)$,
$\alpha(\vec{b})\rightarrow
\alpha_{\mbox{\tiny inel.}}(\vec{b})$;
and in Eq.(\ref{eq20})
$q\rightarrow q_\perp$,
$f(q)\rightarrow 
f_{\mbox{\tiny inel.}}(q_\perp)$,
$\alpha(b)\rightarrow
\alpha_{\mbox{\tiny inel.}}(b)$.
Hereafter, we shall call Eqs.(\ref{eq19}) and (\ref{eq20}) 
with these replacements
Eq.(\ref{eq19}$^\prime$) and Eq.(\ref{eq20}$^\prime$)  respectively.
We note, in order to specify the dependence of 
$f_{\mbox{\tiny inel.}}$
on $\omega^\prime$ and $k_{\|}^\prime$
(that is on 
 $\omega^\prime-\omega$ and $k_{\|}^\prime-k_{\|}$),
further information on
energy-momentum transfer 
in such scattering processes is
needed. This point will be discussed in more detail
in Section \ref{sec:10}.

(B) In scattering processes at large momentum-transfer  
where the magnitude of
$\vec{q}_{\perp}$
is large ($|\vec{q}_{\perp}|^2\gg 0.05\,\mbox{GeV}^2$, say), it
is less probable
to find diffractive scattering
events in which the additional condition 
$|\vec{k}^\prime|=|\vec{k}|$ and $\omega^{\prime}=\omega$
can be satisfied. 
This means, it is expected that most of the diffraction-phenomena
observed in such processes are associated with inelastic diffractive 
scattering.

(C) Change in angle but no change in magnitude of wave-vectors or
frequencies is likely to occur in processes in which neither
absorption nor emission
of light takes place. Hence, it is not difficult to imagine, that the 
above-mentioned condition can be readily satisfied in
cases where the scattering systems are time-independent
macroscopic apertures or objects. But, 
in this connection, we are also forced
to the question:
``How large is the chance for a incident wave {\em not} to change the
magnitude of its wave-vector
in processes in which the scatterers are {\em open, dynamical,
complex systems}, where energy- and momentum-exchange 
take place at anytime and everywhere?!''

The picture for inelastic diffractive scattering
has two basic ingredients:
 
First, having the 
well-known phenomena associated with Frauenhofer's
diffraction
and the properties of 
de Broglie's
matter waves in mind, the beam particles 
($\gamma^\star$, $\gamma$, $\overline{p}$ 
or $p$ shown in Fig.\ref{figure8}) 
in these scattering processes
are considered as high-frequency 
waves passing through a medium. Since, in general, energy- and
momentum-transfer take place during the passage through the medium,
the wave-vector of the outgoing wave differs, in general, from the
incoming one, not only in direction, but also in magnitude. For the
same reason, the frequency 
and the longitudinal component of the wave-vector
of the outgoing wave (that is the energy, and/or the 
invariant mass, as well as the longitudinal momentum
of the outgoing particles) can be different from their
incoming counterparts.

Second, according to the results obtained in Sections
\ref{sec:1}-\ref{sec:7} of this paper, the medium is
a system of color-singlet gluon-clusters ($c_0^\star$'s)
which are in general partly inside and partly outside
the proton --- in form of a ``cluster cloud''. 
Since the average binding energy between such color-singlet aggregates are
of Van der Waals type\cite{Gottfried}, 
and thus it is negligibly small  compared with
the corresponding binding 
energy between colored objects, we expect to see that, 
even at  relatively small values of 
momentum-transfer ($|t|$$<$$1\,\mbox{GeV}^2$,say),
the struck $c_0^\star$ can 
unify with (be absorbed by) the beam-particle, and 
``be carried away'' by the latter,
similar to the process of
``knocking out nucleons'' from nuclear targets in high-energy
hadron-nucleus collisions. It
should, however, be emphasized 
that, in contrast to the nucleons in nucleus, the $c_0^\star$'s
which can exist inside or outside the
confinement-region of the proton
are {\em not} hadron-like (See Sections \ref{sec:3}-\ref{sec:6}
for more details).
They are BTW-avalanches which
have neither a typical size, nor a typical 
lifetime, nor a given static structure. Their 
size- and lifetime-distributions obey simple power-laws
as consequence
of SOC. This means, in the diffraction processes 
discussed here, the size of the scatterer(s), 
and thus the size of the carried-away $c_0^\star$ 
is  in general different in 
every scattering event. 
It should also be emphasized that these
characteristic features
of the scatterer are consequences of
the basic properties of the gluons.

\section{Can such scattering systems be modeled quantitatively?}
\label{sec:10}
To model the proposed picture quantitatively, it is convenient
to consider the scattering system
in the rest frame 
of the proton target. Here, we
choose a right-handed Cartesian coordinate with its origin $O$ at 
the center of the target-proton, 
and the $z$-axis in the direction of the incident 
beam. 
The $xy$-plane in this coordinate system coincides
with the two-dimensional impact-parameter 
space mentioned in connection with Eqs.(\ref{eq19}$^\prime$) and 
(\ref{eq20}$^\prime$)
[which are respectively Eq.(\ref{eq19}) and Eq.(\ref{eq20}) 
after the replacements mentioned in (A) below Eq.(\ref{eq20})],
while the $yz$-plane is the 
scattering plane. We note, since 
we are dealing with inelastic scattering (where the momentum 
transfer,  including its component in the longitudinal direction, 
can be large; in accordance with
the uncertainty principle) it is possible 
to envisage that (the c.m.s.\,of) the incident particle in the beam
meets $c_0^\star$'s at one point $B\equiv(0,b,z)$.
where the projection of $\overline{OB}$ along the $y$-axis characterizes
the corresponding impact parameter 
$\vec b$.
We recall that such $c_0^\star$'s are avalanches 
initiated by local perturbations (caused by
local gluon-interactions associated with 
absorption or emission of one or more gluons;
see Sections \ref{sec:1} - \ref{sec:7} for details) of SOC states 
in systems of interacting soft gluons. Since gluons 
carry color, the interactions which lead to the formation of
{\em color-singlet} gluon-clusters ($c_0^\star$) must take place 
inside the confinement region of  the proton. This means, while a 
considerable part of such $c_0^\star$'s in the cloud  can be outside
the proton, the location $A$, where such an avalanche is initiated, 
{\em must} be {\em inside} the proton. That is, in terms of 
$\overline{OA}\equiv r$, $\overline{AB}\equiv R_A(b)$, and  
proton's radius $r_p$,
we have $r\le r_p$ and 
$[R_A(b)]^2=b^2+z^2+r^2-2(b^2+z^2)^{1/2} r \cos\angle BOA$. 
For a given impact parameter
$b$, it is useful to know the distance $R_A(b)$ between $B$ and $A$,
as well as  ``the average squared distance''
$\langle R_A^2(b)\rangle =  b^2+z^2 + a^2$,
$a^2\equiv 3/5\,r_p^2$,
which is
obtained by averaging  over all allowed locations of $A$
in the confinement region.
That is, we can model {\em the effect of confinement} in 
cluster-formation by picturing that all the avalanches,
in particular those
which contribute to 
scattering events characterized by a given $b$ and a given $z$ 
are initiated from an ``effective initial point'' $\langle
A_B\rangle$,
because only the mean distance between $A$ and $B$ 
plays a role.
(We note, since we are dealing with a complex system with many degrees 
of freedom, in which
$B$ as well as $A$ are randomly chosen points in
space, 
we can compare {\em the mean distance} between $B$ and $A$ 
with the mean free path in a gas mixture
of two kinds of gas molecules ---
``Species $B$'' and ``Species $A$'' say, where those of the latter kind are
confined inside a subspace called ``region $p$''.
For {\em a given mean distance}, and {\em a given point $B$},
there is in general a set of $A$'s inside the ``region $p$'',
such that their distance to $B$ is {\em equal} to the given mean
value. Hence it is useful to introduce a {\em representative} point
$\langle A_B\rangle$, such that the distance
between $\langle A_B\rangle$ and $B$ is equal 
to the given mean distance.)
Furthermore, since an avalanche is a dynamical object, it may propagate 
within its lifetime
in any one of the $4\pi$
directions away from $\langle A_B\rangle$. (Note: avalanches of
the same size may have different lifetimes, different structures, 
as well as 
different shapes. The location of an avalanche
in space-time is referred to its center-of-mass.)
Having seen how SOC and confinement can be implemented in describing
the properties and the dynamics of the $c_0^\star$'s,
which are nothing else but BTW-avalanches in systems of interacting
soft gluons,
let us now go one step further, and discuss how these
results can be used to obtain the amplitudes in impact-parameter-space 
that leads, via
Eq.(\ref{eq20}), to the scattering amplitudes.

In contrast to the usual cases, where the scatterer  
in the optical
geometrical picture of a diffractive scattering process
is an aperture, or an 
object, with a given static structure,
the scatterer in the proposed picture is
an open, dynamical, complex system of $c_0^\star$'s.
This implies in particular: The object(s), which the beam particle 
hits, has (have) neither a typical size, nor a typical lifetime, nor a
given static structure.

\begin{figure}
\psfig{figure=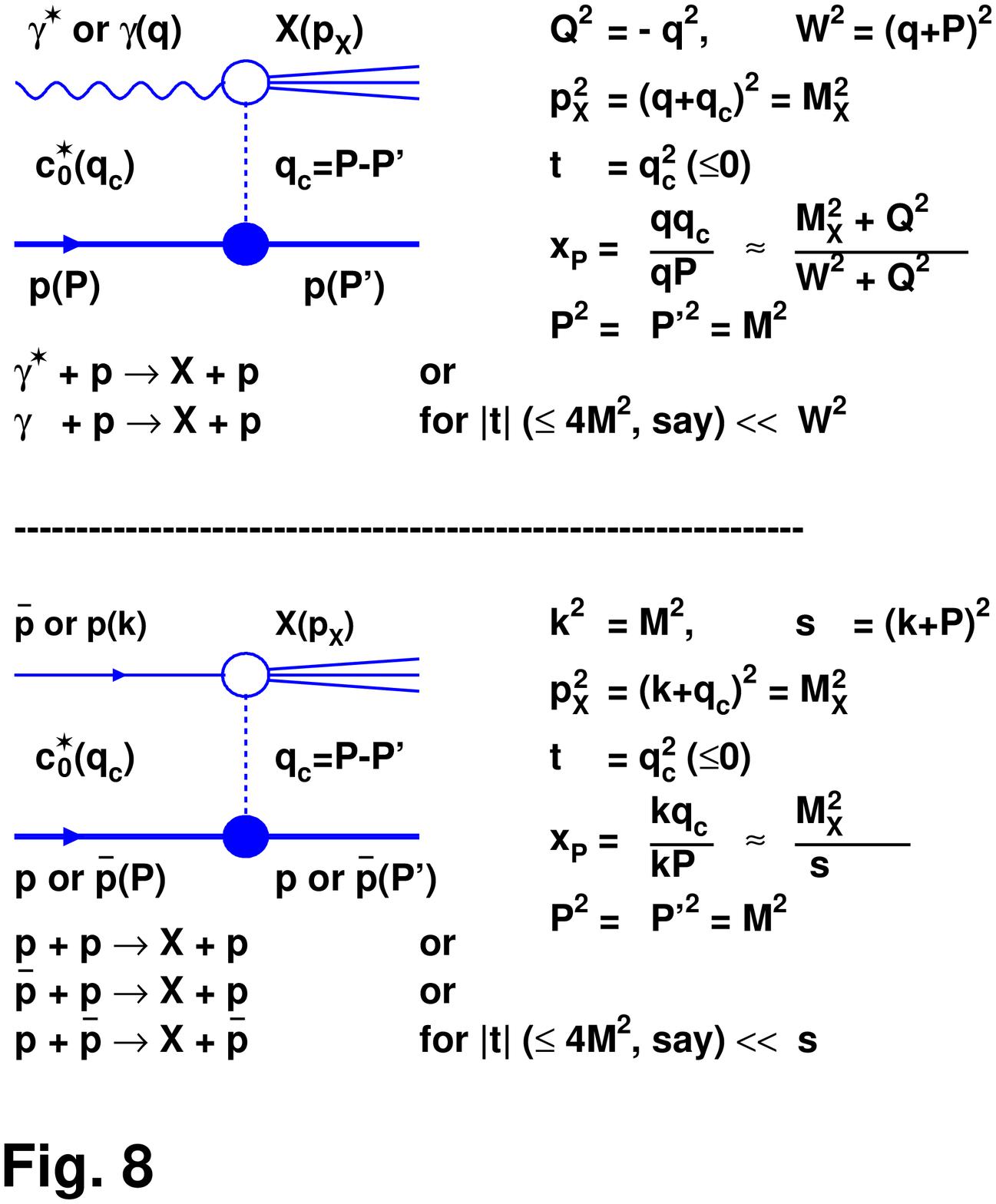,width=16.0cm}
\caption{\label{figure8}
        Diagrams for different single diffractive 
        reactions, together with the definitions of the relevant
        kinematic variables.}
\end{figure}

\begin{figure}
\psfig{figure=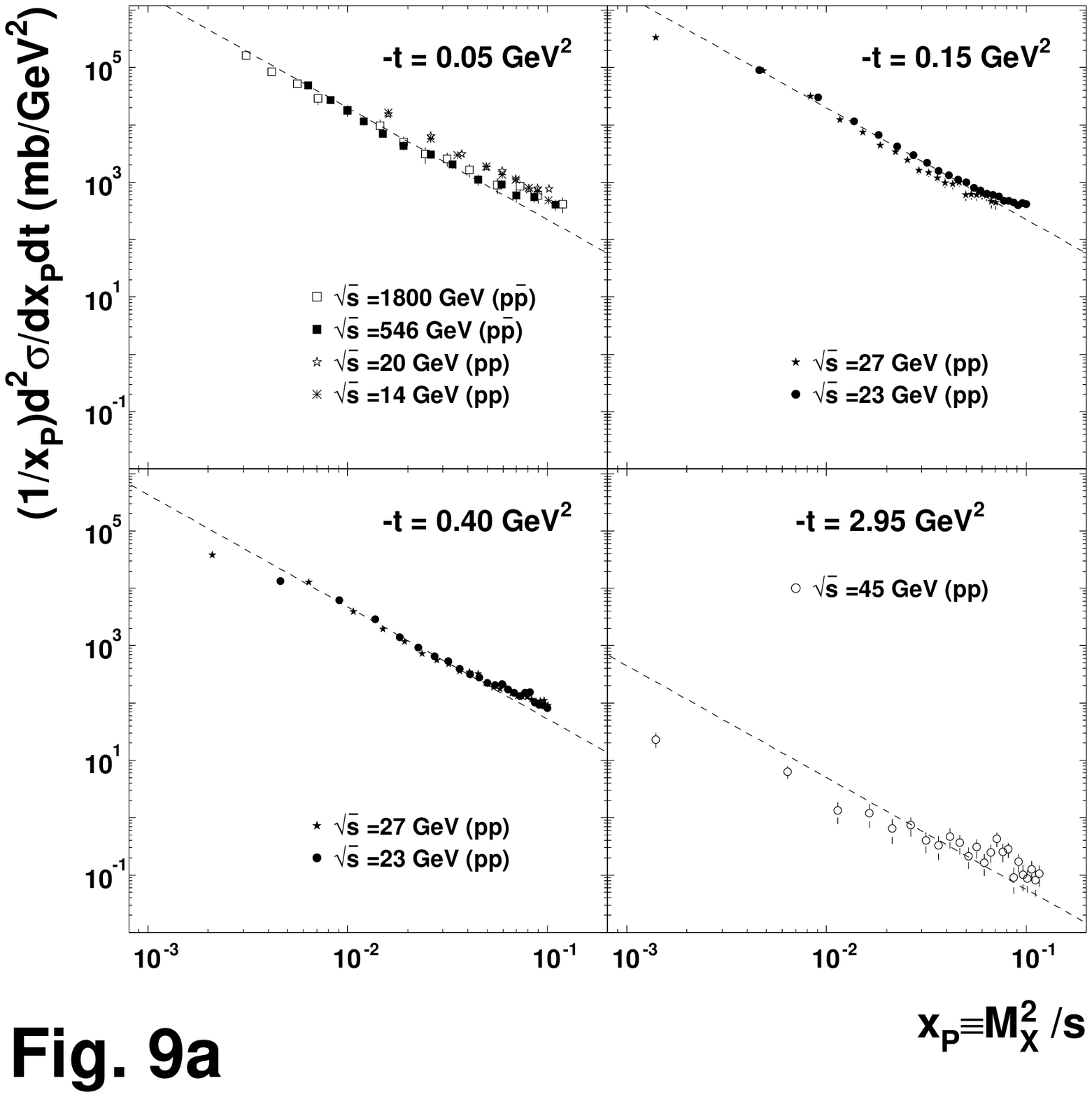,width=16.0cm}
\psfig{figure=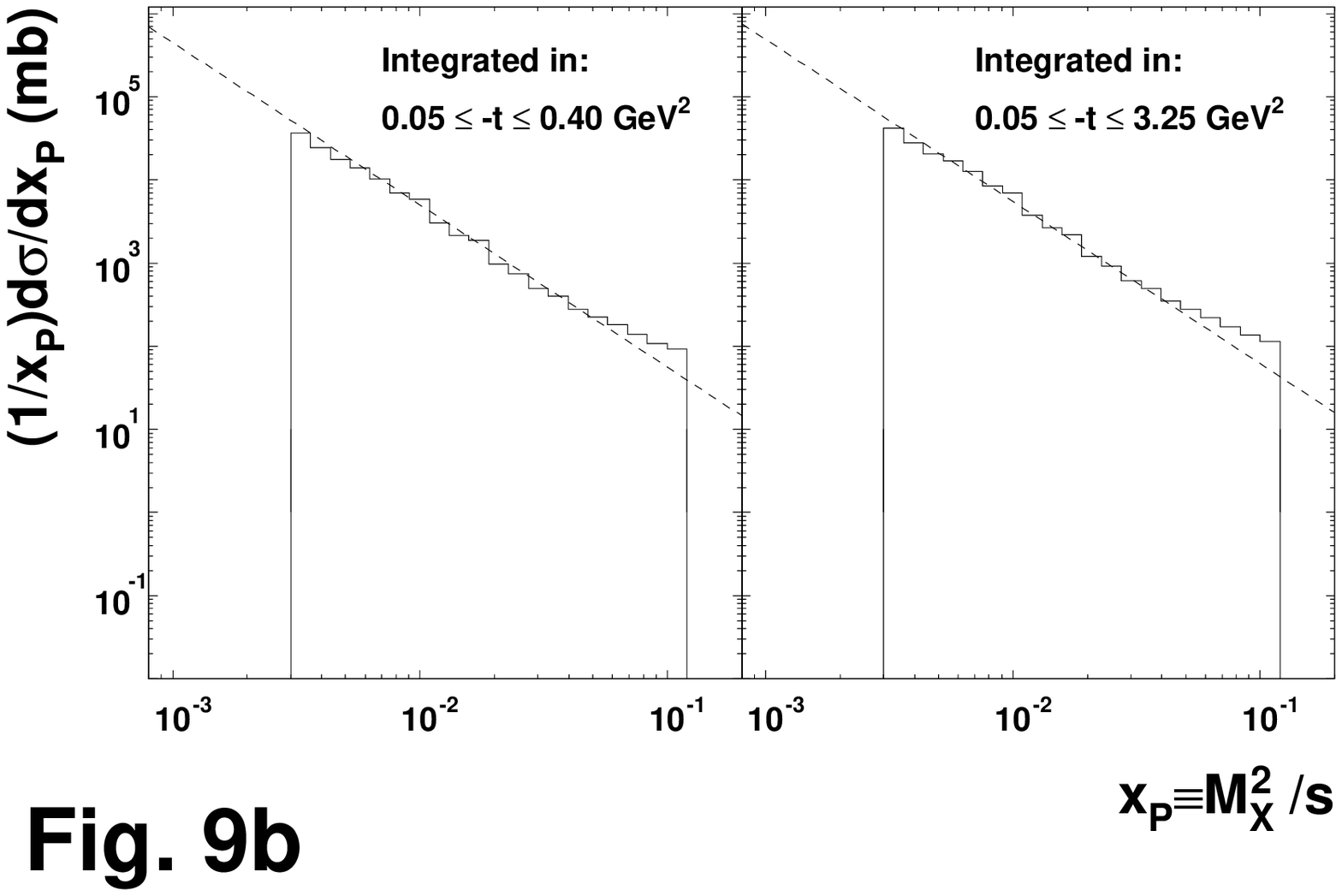,width=16.0cm}
\caption{\label{figure9}
         a) $(1/x_P)d^2\sigma/dx_P dt$ 
           for  single diffractive 
           $p + p \rightarrow p + X$ and 
           $p + \bar{p} \rightarrow p + X$  reactions 
           is plotted as a function of $x_P$ 
           at different values of $t$ and $\sqrt{s}$.
           The data are taken from Ref.[\ref{r4},\ref{r5}].
         b) The integrated (with respect to two different
            $|t|$-ranges) differential cross section
            $(1/x_P)d\sigma/dx_P$ 
           for  single diffractive 
           $p + p \rightarrow p + X$ and 
           $p + \bar{p} \rightarrow p + X$  reactions 
           is plotted as a function of $x_P$.}
\end{figure}

With these in mind, let us now come back to our discussion on the
double differential cross section $d^2\sigma/dt\,d(M_x^2/s)$.
Here, we need to determine
the corresponding amplitude 
$\alpha_{\mbox{\tiny inel.}}(b)$
in Eq.(\ref{eq20}$^\prime$) [see the discussion in (A) below 
Eq.(\ref{eq20})]. What we wish to do now, is to 
focus our attention on those
scattered matter-waves whose de Broglie wavelengths are
determined by the energy-momentum of the scattered 
object, whose invariant mass  is $M_x$.
For this purpose, we characterize
the corresponding $\alpha_{\mbox{\tiny inel.}}(b)$ 
by considering it as a function of
$M_x$, or $M_x^2/s$, or $x_P$. We 
recall in this connection that, for inelastic diffractive 
scattering processes in hadron-hadron collisions, the quantity $M_x^2/s$
is approximately equal to $x_P$, which is 
the momentum fraction carried by the struck
$c_0^\star$'s with respect  to the incident beam
(see Fig.\ref{figure8} for more details; note however
that $q_c$, $k$ and $p_x$ in Fig.\ref{figure8} correspond respectively to
$q$, $k$ and $k^\prime$ in the discussions here.).
Hence, we shall write hereafter 
$\alpha(b\,|M_x^2/s)$ or $\alpha(b\,|x_P)$
instead of the general expression
$\alpha_{\mbox{\tiny inel.}}(b)$.
This,
together with Eq.(\ref{eq20}$^\prime$), leads to the corresponding
scattering amplitude
$f(q_\bot |x_P)$, and 
thus to the corresponding double differential cross-section
$d^2\sigma/dt\,dx_P$,
in terms of the variables
$|t|\approx |\vec{q}_{\bot}|^2$ 
and $x_P\approx M_x^2/s$ in the kinematical region: $|t|\ll M_x^2 \ll s$.

\section{The role played by the space-time properties of the
gluon-clusters}
\label{sec:11}
For the determination of $\alpha(b\,|x_P)$, it is of considerable
importance to recall the following space-time properties of the
$c_0^\star$'s which are BTW-avalanches due to SOC:

(i) SOC dictates, that there are BTW-avalanches of all sizes 
(which we denote by different $S$ values), and
that the probability amplitude of finding an avalanche of size
$S$ can be obtained from the size-distribution
$D_S(S)=S^{-\mu}$ where the experimental results
presented in Sections \ref{sec:1} - \ref{sec:7}
 show: $\mu\approx 2$.
This means, $D_S(S)$ contributes a factor 
$S^{-1}$, thus a factor 
$x_P^{-1}$ to the scattering amplitude $\alpha(b\,|x_P)$.
Here, as well as in (ii), we take into account
(see Sections \ref{sec:4} and \ref{sec:5} for details) 
that the size
$S$ of a $c_0^\star$ is directly proportional to the
total amount of the energy the $c_0^\star$ carries;
the amount of energy is $x_P P^0$,
where $P^0$ is the total energy of the proton, and $x_P$ is the 
energy fraction carried by the $c_0^\star$.

(ii) QCD implies\cite{Gottfried} that the 
interactions between two arbitrarily
chosen colored constituents of $c_0^\star$ is stronger than those 
between two $c_0^\star$'s,
because the latter should be interactions of Van der
Waals type. This means, the struck $c_0^\star$ can 
unify with the beam-particle (maybe by absorbing each other), and
viewed from any Lorentz frame in which the beam-particle has a larger
momentum than the $c_0^\star$, the latter is
``carried away'' by the beam particle.
Geometrically, the chance for the 
beam-particle to hit an $c_0^\star$
of size $S$ ( on the plane perpendicular to the incident axis) 
is proportional to  the area that can be struck by the (c.m.s.) of the 
beam particle. The area is the $2/3$-power of the volume $S$,
$S^{2/3}$,
and thus it is proportional to $x_P^{2/3}$.

(iii) Based on the above-mentioned picture 
in which the $c_0^\star$'s propagate isotropically
from $\langle A_B\rangle$,  the
relative number-densities at different $b$-values can be readily 
evaluated. Since for a given $b$, the distance
in space between $\langle A_B\rangle$ and
$B\equiv (0,b,z)$ is simply 
$(b^2+z^2+a^2)^{1/2}$, the number of $c_0^\star$'s which pass a unit area 
on the shell of radius $(b^2+z^2+a^2)^{1/2}$ centered at $\langle A_B\rangle$
is proportional to $(b^2+z^2+a^2)^{-1}$, provided that (because of causality)
the lifetimes ($T$'s) of these $c_0^\star$'s are not shorter than
$\tau_{\mbox{\tiny min}}(b)$. The latter is the time interval for a
$c_0^\star$ to travel from  $\langle A_B\rangle$ to $B$.
This means, because of the space-time properties of such $c_0^\star$'s, it 
is of considerable importance to note: 
First, only $c_0^\star$'s having lifetimes
$T\ge\tau_{\mbox{\tiny min}}(b)$ can contribute to such a collision
event. Second, during the propagation from $\langle A_B\rangle$ to
$B$, the motion of such a $c_0^\star$ has to be considered as
Brownian. In fact, the continual, and more or less random, impacts
received from the neighboring objects on its path leads us to the well 
known \cite{StatMech} result that the time elapsed is proportional to
the mean-square displacement. 
That is: $\tau_{\mbox{\tiny min}}(b)\propto b^2+z^2+a^2$.
Furthermore, we recall that $c_0^\star$'s are due to SOC, 
and thus the chance for a $c_0^\star$ of lifetime  
$T$ to exist is $D_T(T) \propto T^{-\nu}$ 
where the experimental value (see Sections \ref{sec:1} - \ref{sec:7}) for
$\nu$ is $\nu\approx 2$. Hence, by integrating 
$ T^{-2}$ over $T$ from $\tau_{\mbox{\tiny min}}(b)$ to infinity,
we obtain
the fraction associated with all those whose lifetimes satisfy $T\ge
\tau_{\mbox{\tiny min}}(b)$: This fraction 
is $\tau_{\mbox{\tiny min}}(b)^{-1}$, and thus
a constant times  $(b^2+z^2+a^2)^{-1}$.

\begin{figure}
\psfig{figure=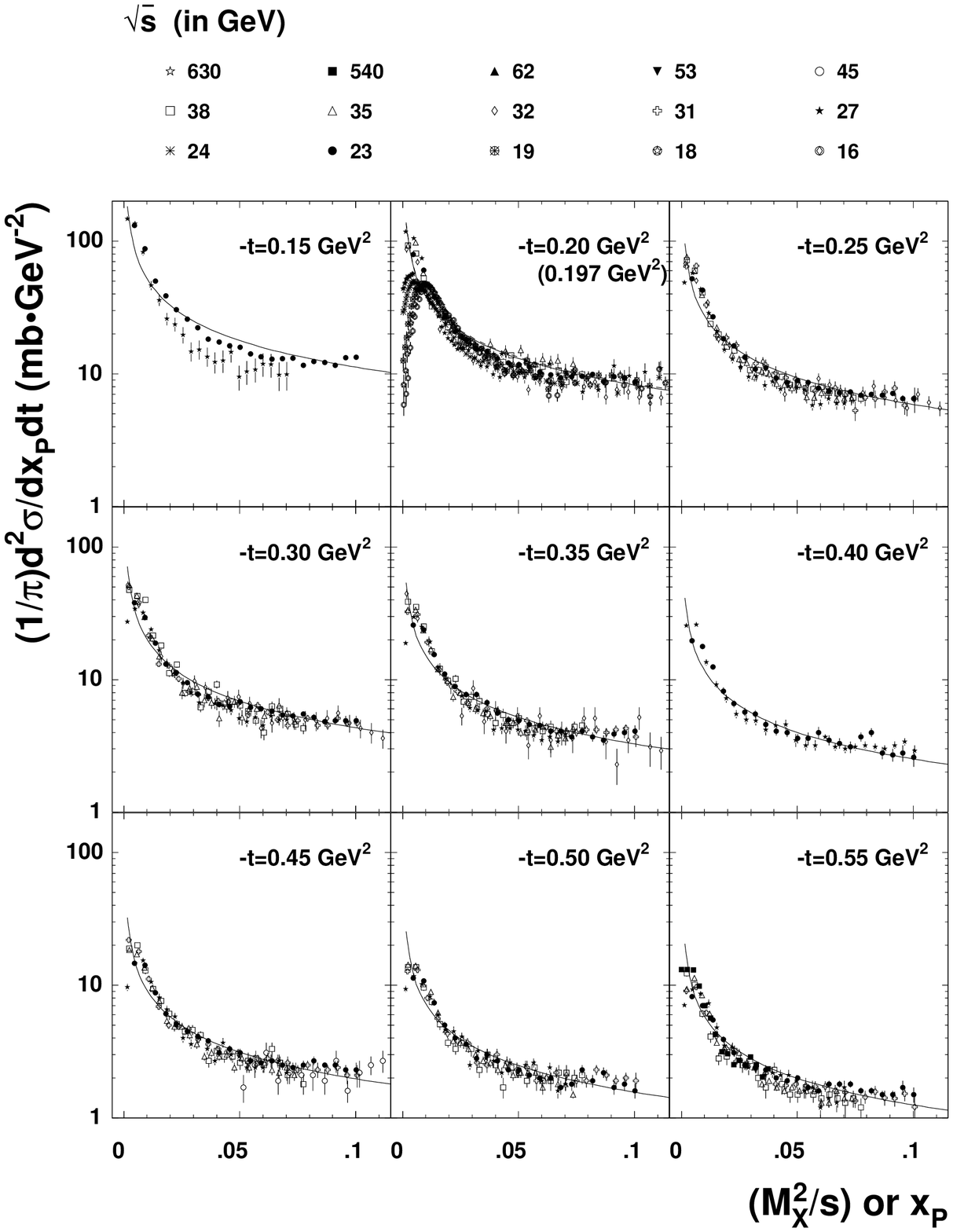,width=16.0cm}
\psfig{figure=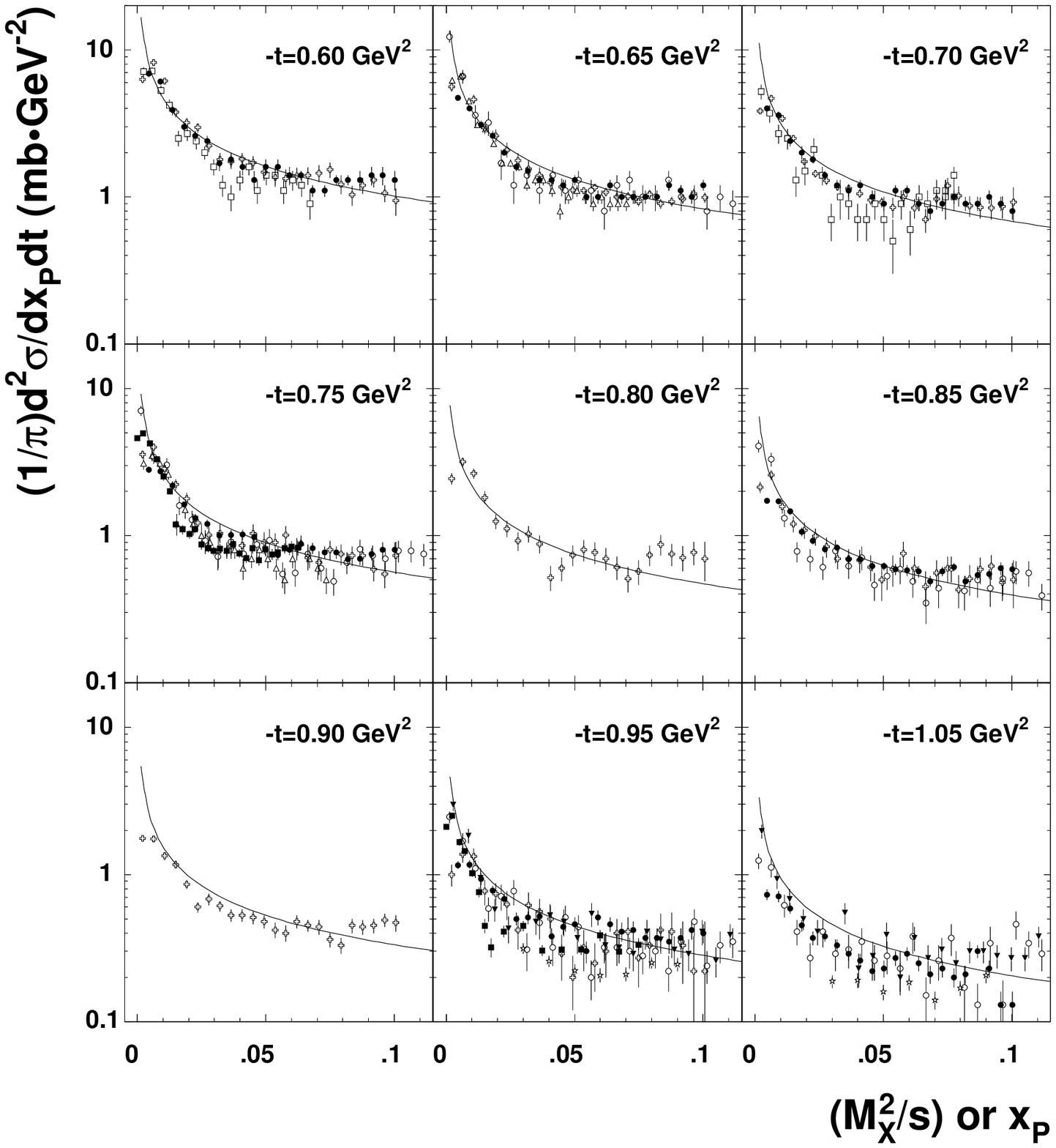,width=16.0cm}
\psfig{figure=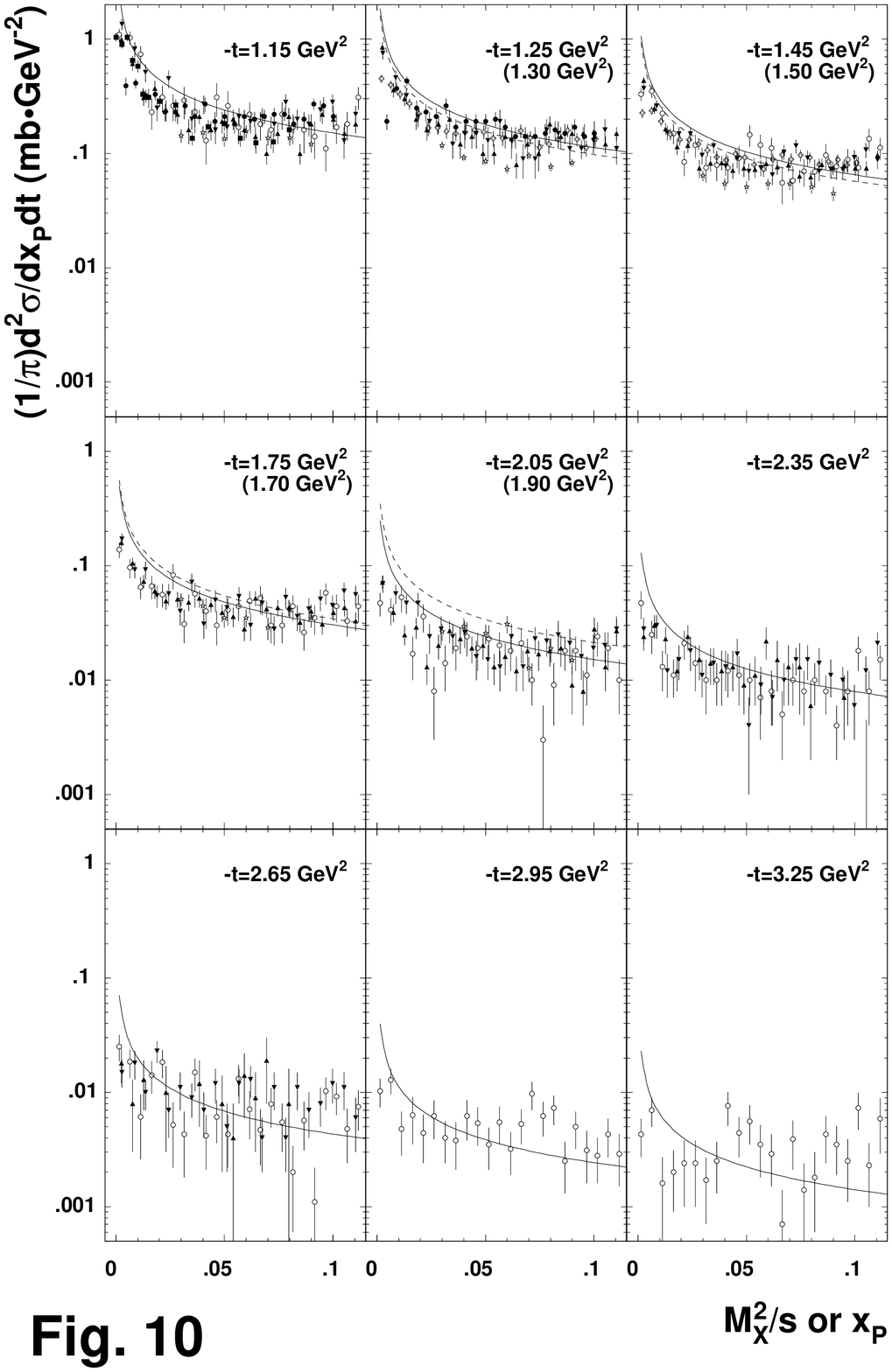,width=16.0cm}
\caption{\label{figure10}
The double differential cross section $(1/\pi)\,d^2\sigma/ dt\,d(M_x^2/s)$ 
for single diffractive $pp$ and $\bar{p}p$ reactions
is shown as function of
 $x_P$ at fixed values of $t$ where
$0.15$$\mbox{\,GeV}^2$$\le$$|t|$$\le$$3.25$$\mbox{\,GeV}^2$.
The data are taken from Refs.
[\ref{r4},\ref{ppbardata}]. The solid curve is the
result obtained from Eq.(\ref{eq24}). 
The dashed curve stands for the result obtained from the same formula
by using the $t$-value given in the bracket.}
\end{figure}

The amplitude $\alpha(b\,|x_P)$ can now be obtained from
the probability amplitude for $c_0^\star$-creation
mentioned in (i), by 
taking the weighting factors mentioned in
(ii) and (iii) into account, and by integrating\cite{zInt} over $z$ . 
The result is:
\begin{eqnarray}
\label{eq21}
\alpha(b\,|x_P) = \mbox{const.} x_P^{-1/3}(b^2+a^2)^{-3/2}\mbox{\,.}
\end{eqnarray}
By inserting this
probability amplitude in
impact-parameter-space, for the beam particle to encounter
a $c_0^\star$, which carries a
fraction $x_P$ of the proton's total energy, in  
Eq.(\ref{eq20}$^\prime$) [which is Eq.(\ref{eq20}) 
with the following replacements:
$q\rightarrow q_\perp$,
$f(q)\rightarrow f_{\mbox{\tiny inel.}}(q|x_P)$ and
$\alpha(b)\rightarrow \alpha_{\mbox{\tiny inel.}}(b|x_P)$]
we obtain
the
corresponding probability amplitude $f(q|x_P)$ in 
momentum-space:
\begin{equation}
\label{eq22}
f(q_\bot |x_P)=\mbox{const.}\int_0^{\infty} 
b\, db\,x_P^{-1/3}(b^2+a^2)^{-3/2}J_0(q_\bot b),
\end{equation}
where $q_\bot =|\vec{q}_\bot |\approx \sqrt{|t|}$ 
(in the small $x_P$-region, $x_P<0.1$, say)
is the 
corresponding momentum-transfer.
The integration can be carried out analytically\cite{Integrals},
and the result is 
\begin{equation}
\label{eq23}
f(q_\bot |x_P)={\rm const.}x_P^{-1/3}\exp (-a q_\bot )\mbox{\,.}
\end{equation}
Hence, the corresponding double differential cross-section
$d^2\sigma/ dt dx_P$ can approximately be written as
\begin{equation}
\label{eq24}
\frac{1}{\pi}{d^2\sigma\over dt\,dx_P}=N x_P^{-2/3}\exp (-2a\sqrt{|t|}),
\end{equation}
where $N$ is an unknown normalization constant. Because of the
kinematical relationship $x_P\approx M_x^2/s$ for 
single diffractive scattering in proton-proton and
proton-antiproton collisions
(see Fig.\ref{figure8} for more details),
this can be, and
should be, compared with the measured double differential
cross-sections $d^2\sigma/dt\,d(M_x^2/s)$ at different $t$- and
$s$-values and for different missing masses $M_x$
in the region $M_x^2/s\ll 1$ 
where $q_\perp$ is approximately $\sqrt{|t|}$. 
The comparison is shown
in Fig.\ref{figure10}. 
Here, we made use of the fact that $a^2\equiv 3/5\,r_p^2$, 
where $r_p$ is the proton radius, and calculated $a$ by setting $r_p^2$
to be the 
well-known \cite{Halzen} 
mean square proton charge radius, the value of which is
$r_p^2 = (0.81\,\mbox{fm})^2$.
The result we obtained is:
$a= 3.2\,\mbox{GeV}^{-1}$. The unknown normalization constant
is determined by inserting this calculated value for $a$ in 
Eq.(\ref{eq24}), and 
by comparing the right-hand-side of this equation with
the $d^2\sigma/dt\,d(M_x^2/s)$ data taken
at $|t|= 0.2\,\mbox{GeV}^2$. The value is $N=31.1\,\mbox{mb
GeV}^{-2}$. All the curves shown in Fig.\ref{figure10} 
are obtained by inserting these 
values for $a$ and $N$ in Eq.(\ref{eq24}).

While the quality of the obtained result, namely the expression given
on the right-hand-side of Eq.(\ref{eq24})
together with the above-mentioned values for $a$ and $N$,
can be readily judged by comparing
it with the data, or by counting the unknown parameters, or both, it
seems worthwhile to recall the following:
The two basic ingredients of the proposed picture which have been used 
to derive this simple analytical expression are: first, the well-known 
optical analogy, and second, the properties of the dynamical
scattering system. The latter is what we have learned through the
data-analysis presented in Sections \ref{sec:1} - \ref{sec:7}.

Based on the theoretical arguments and experimental indications 
for the observation (see Ref.[\ref{r10a}] and Section \ref{sec:1} 
- \ref{sec:7} of this paper
for details) that the characteristic features of
inelastic diffractive scattering processes are approximately
independent of the incident energy and independent of the
quantum-numbers of the beam-particles, the following
results are expected:
The explicit formula for the double differential cross-section 
as shown in Eq.(\ref{eq24}) should also be valid for the reactions 
$\gamma p\rightarrow X p$ and 
$\gamma^\star p\rightarrow X p$.
While the normalization constant $N$ (which should in particular
depend on the geometry of the beam particle) is expected to be
different for different reactions, everything else --
especially the ``slope'' as well as the power of $x_P$ should be
exactly the same as in $pp$- and $p\bar{p}$-collisions.
In this sense, Eq.(\ref{eq24}) 
with $a^2=3/5r_P^2$ ($r_P$ is the proton radius) 
is our prediction for $\gamma p \rightarrow X p$ and $\gamma^\star p
\rightarrow X p$ which can be measured at HERA.

\begin{figure}
\psfig{figure=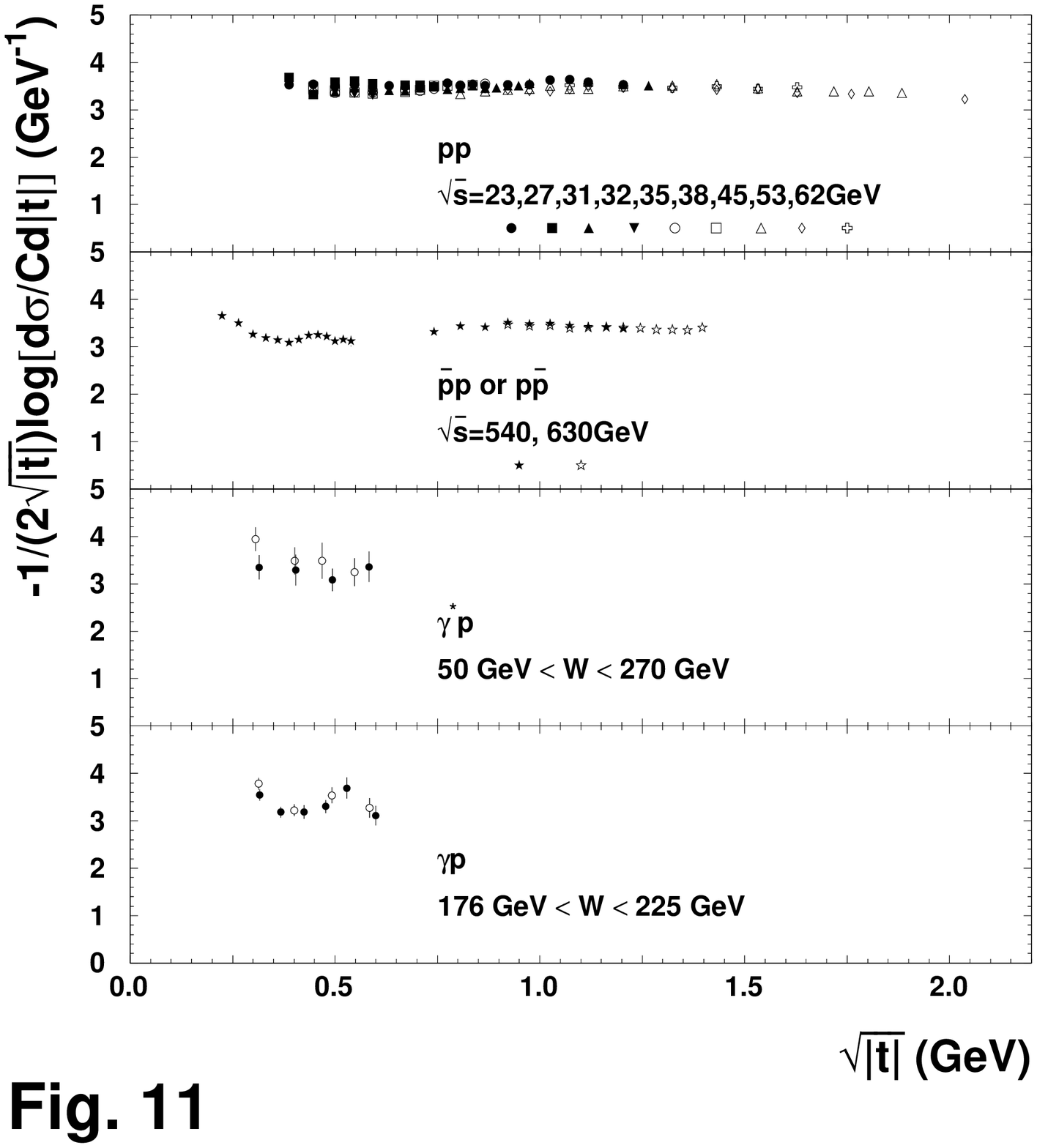,width=16.0cm}
\caption{\label{figure11}
The quantity $(-1/(2 \sqrt{|t|})) \log{[\frac{1}{C}\,d\sigma/ dt]}$ 
is plotted versus $\sqrt{|t|}$ for different single diffractive
reactions
in the range
$0.2$$\mbox{\,GeV}^2$$\le$$|t|$$\le$$4$$\mbox{\,GeV}^2$.
The data are taken from Refs. 
[\ref{r4},\ref{ppbardata},\ref{dsigma/dt}].
Here, $C$, the normalization constant is first
determined by performing a two-parameter fit of 
the corresponding $d\sigma/dt$-data to Eq.(\ref{eq25}).}
\end{figure}

Furthermore, in order to obtain the integrated differential cross-section
$d\sigma/dt$, which has also been measured for different
reactions at different incident energies, we only need to
sum/integrate over $x_P$ in the given kinematic range ($x_P<0.1$,
say). The result is
\begin{equation}
\label{eq25}
{d\sigma\over dt}(t)=C\exp (-2a\sqrt{|t|}),
\end{equation}
where $C$ is an unknown normalization
constant. 
While this observation
has already been briefly discussed in the
previous Letter \cite{r10a}, 
we now show the result of a further 
test of its universality: In Fig.\ref{figure11}, we plot
\begin{equation}
\label{eq26}
-{1\over 2\sqrt{|t|}}\log [\frac{1}{C}\,{d\sigma\over dt}(t)] \mbox{ vs. } t 
\end{equation}
{\em for different reactions at different incident energies} 
in the range 
$0.2$$\mbox{\,GeV}^2$$\le$$|t|$$\le 4\mbox{\,GeV}^2$.
Here we see in particular that, 
measurements of $d\sigma/dt$ for $\gamma^{\star}p$ and
$\gamma p$ reactions at larger $|t|$-values would be very useful.

\section{Concluding remarks}
\label{sec:12}
Based on the characteristic properties of the gluons ---
in particular the local gluon-gluon coupling prescribed by
the QCD-Lagrangian, the confinement, and the non-conservation 
of gluon-numbers, we suggest that a system of interacting
soft gluons should be considered as an open dynamical complex system
which is in general far away from equilibrium.
Taken together with the observations made 
by Bak, Tang and Wiesenfeld (BTW) \cite{r11,r12},
we are led to the conclusion, 
that self-organized criticality (SOC) and 
thus BTW-avalanches exist in such systems, and that 
such avalanches 
manifest themselves in form of color-singlet gluon-clusters ($c_0^\star$) 
in inelastic diffractive scattering processes.

In order to test this proposal, we performed a systematic
data-analysis, the result of which is presented in Sections
\ref{sec:1} - \ref{sec:7}: It is shown that the size-distributions, 
and the lifetime-distributions, of
such $c_0^\star$'s {\em indeed} exhibit power-law behaviors which are known 
as the fingerprints of SOC \cite{r11,r12}. 
Furthermore, it is found that such exponents
are approximately the same for different reactions and/or at different
incident energies --- indicating the expected universality and
robustness of SOC. Hence, the following picture emerges:
For the beam particle (which may be a 
virtual photon, or a real photon, or 
a proton, or an antiproton; see Fig.\ref{figure8} for more details) 
in an inelastic diffractive scattering process 
off proton (one may wish to view this from a ``fast moving frame''
such as the c.m.s. frame), 
the target proton appears as a cloud of $c_0^\star$'s
which exist inside and outside the confinement
region of the proton. The size-distribution $D_S(S)$ and the
lifetime-distribution $D_T(T)$ can be expressed as $S^{-\mu}$ and
$T^{-\nu}$ respectively, where 
the empirical values for $\mu$ and $\nu$ are
$\mu$$\approx$$\nu$$\approx$$2$,
independent of the incident energy, and independent of the quantum
numbers of the beam particles.

What do we learn from this? Is this knowledge
helpful in understanding hadronic structure and/or
hadronic reactions 
in Particle Physics? In particular, can 
this knowledge be used to do {\em quantitative} 
calculations --- especially those, the results of
which could {\em not} be
achieved otherwise?

In order to demonstrate how the obtained knowledge can be used to 
relate hadron-structure and hadronic reactions in general, and to
perform quantitative calculations in particular, we discuss the
following question --- a question which has been with the high-energy 
physics community for many years:

``Can the measured double differential cross section 
$d^2\sigma/dt\,d(M_x^2/s)$ 
for 
inelastic diffractive scattering in
proton-proton and in antiproton-proton collisions,
in the kinematical region given by 
$0.2\mbox{\,GeV}^2$$\le$$|t|$$\le$ $3.25 \mbox{\,GeV}^2$, 
$16\mbox{\,GeV}$$\le$$\sqrt{s}$$\le$ $630\mbox{\,GeV}$,
and $M_x^2/s$ $\le$$0.1$, be understood in terms of optical
geometrical concepts?'' 

The answer to this question is ``Yes!'', and the 
details are presented in Sections \ref{sec:9} - \ref{sec:11} where
the following is explicitly shown:
The characteristic features of the existing
$d^2\sigma/dt\,d(M_x^2/s)$-data are very much the same as those in
optical diffraction, provided that the high-energy beams are
considered as high-frequency waves, and the scatterer is a
system of color-singlet gluon-clusters ($c_0^\star$) 
described in Sections \ref{sec:1} - \ref{sec:7} of this paper.
Further measurements of double differential cross sections, especially
in $\gamma^{\star}p$- and $\gamma p$-reactions, will be helpful in
testing the ideas presented here.

\section*{Acknowledgement}
We  thank  P. Bak, X. Cai, D. H. E. Gross, C. S. Lam, Z. Liang,
K. D. Schotte, C. B. Yang, E. Yen, and W. Zhu
for helpful discussions, R. C. Hwa, C. S. Lam and J. Pan for
correspondence, and FNK der FU-Berlin
for financial support.
Y. Zhang also thanks Alexander 
von Humboldt Stiftung for the fellowship granted to him.

\newpage

\begin{thebibliography}{99}
\bibitem{zInt}
\label{zInt}
To be more precise, by 
taking all the mentioned factors into account we obtain
$\alpha(b,z|x_P)\propto \mbox{const }
x_P^{-1+2/3} (b^2+z^2+a^2)^{-1-1}$. Note also that $\int_{-\infty}^\infty 
(b^2+z^2+a^2)^{-2} dz \propto (b^2+a^2)^{-3/2}$.
\end{thebibliography}
\end{document}